 \documentclass[twocolumn]{aastex631}
\usepackage{footnote}

\newcommand{\mum}{{\hbox {\,$\mu$m}}}

\newcommand{\Lsun}{{\hbox {$L_\odot$}}}

\usepackage{amsmath}

\def\aco  {\ifmmode{{\rm CO}(1$-$0)}\else{CO(1$-$0)}\fi}
\def\bco  {\ifmmode{{\rm CO}(2$-$1)}\else{${\rm CO}(2-1)$}\fi}
\def\cco   {\ifmmode{{\rm CO}(3$-$2)}\else{${\rm CO}(3-2)$}\fi}
\def\dco  {\ifmmode{{\rm CO}(4$-$3)}\else{${\rm CO}(4-3)$}\fi}
\def\eco  {\ifmmode{{\rm CO}(5$-$4)}\else{${\rm CO}(5-4)$}\fi}

\shorttitle{\aco\ in $z=2-4$ luminous DSFGs}
\shortauthors{Stanley et al.}
\begin{document}

\title{Resolved CO(1$-$0) emission and gas properties in  
luminous dusty star forming galaxies at $z=2-4$}

\author[0000-0003-2156-8386]{F. Stanley}
\affiliation{Sorbonne Universit\'e, UPMC Universit\'e Paris 6 \& CNRS, UMR 7095, Institut d'Astrophysique de Paris, 98b Boulevard Arago, 75014 Paris, France}
\affiliation{Institut de Radioastronomie Millim\'etrique (IRAM), 300 Rue de la Piscine, 38400 Saint-Martin-d'H\`eres, France}
\author[0000-0002-0675-0078]{B.\,M. Jones}
\affiliation{I. Physikalisches Institut, Universit\"at zu K\"oln, Z\"ulpicher Strasse 77, D-50937 K\"oln, Germany}
\affiliation{Jodrell Bank Centre for Astrophysics, Department of Physics and Astronomy, School of Natural Sciences, The University of Manchester, Manchester, M13 9PL, UK}
\author[0000-0001-9585-1462]{D.\,A. Riechers}
\affiliation{I. Physikalisches Institut, Universit\"at zu K\"oln, Z\"ulpicher Strasse 77, D-50937 K\"oln, Germany}
\author[0000-0002-8112-9991]{C. Yang}
\affiliation{Department of Space, Earth and Environment, Chalmers University of Technology, Onsala Space Observatory, SE-439 92 Onsala, Sweden}
\author[0000-0002-0320-1532]{S. Berta}
\affiliation{Institut de Radioastronomie Millim\'etrique (IRAM), 300 Rue de la Piscine, 38400 Saint-Martin-d'H\`eres, France}
\author[0000-0003-2027-8221]{P. Cox}
\affiliation{Sorbonne Universit\'e, UPMC Universit\'e Paris 6 \& CNRS, UMR 7095, Institut d'Astrophysique de Paris, 98b Boulevard Arago, 75014 Paris, France}
\author[0000-0002-5268-2221]{T.\,J.\,L.\,C. Bakx}
\affiliation{Division of Particle and Astrophysical Science, Graduate School of Science, Nagoya University, Aichi 464-8602, Japan}
\affiliation{National Astronomical Observatory of Japan, 2-21-1, Osawa, Mitaka, Tokyo 181-8588, Japan}
\author{A. Cooray}
\affiliation{Department of Physics and Astronomy, University of California, Irvine, CA92697, USA}
\author[0000-0001-7147-3575]{H. Dannerbauer}
\affiliation{Instituto de Astrof\'isica de Canarias (IAC), E-38205 La Laguna, Tenerife, Spain}
\affiliation{Universidad de La Laguna, Dpto. Astrof\'isica, E-38206 La Laguna, Tenerife, Spain}
\author[0000-0002-1318-8343]{S. Dye}
\affiliation{School of Physics and Astronomy, University of Nottingham, University Park, Nottingham, NG7 2RD, UK}
\author{D.\,H. Hughes}
\affiliation{Instituto Nacional de Astrof\'isica, \'Optica y Electr\'onica, Luis Enrique Erro 1, Santa Mar\'ia Tonantzintla, Puebla 72840, Mexico}
\author[0000-0001-5118-1313]{R.\,J. Ivison}
\affiliation{European Southern Observatory, Karl-Schwarzschild-Strasse 2, 85748 Garching, Germany}
\author[0000-0002-8412-7951]{S. Jin}
\affiliation{Cosmic Dawn Center (DAWN)}
\affiliation{DTU Space, Technical University of Denmark, Elektrovej 327, DK-2800 Kgs. Lyngby, Denmark}
\author[0000-0003-1939-5885]{M. Lehnert}
\affiliation{Centre de Recherche Astrophysique de Lyon - CRAL, CNRS UMR 5574, UCBL1, ENSLyon, 9 avenue Charles Andr\'e, F-69230 Saint-Genis-Laval}
\author[0000-0002-7176-4046]{R. Neri}
\affiliation{Institut de Radioastronomie Millim\'etrique (IRAM), 300 Rue de la Piscine, 38400 Saint-Martin-d'H\`eres, France}
\author[0000-0002-4721-3922]{A. Omont}
\affiliation{Sorbonne Universit\'e, UPMC Universit\'e Paris 6 \& CNRS, UMR 7095, Institut d'Astrophysique de Paris, 98b Boulevard Arago, 75014 Paris, France}
\author[0000-0001-5434-5942]{P. van der Werf}
\affiliation{Leiden Observatory, Leiden University, P.O. Box 9513, 2300 RA Leiden, The Netherlands}
\author[0000-0003-4678-3939]{A. Wei{\ss}}
\affiliation{Max-Planck-Institut f\"ur Radioastronomie, Auf dem H\"ugel 69, 53121 Bonn, Germany}

\begin{abstract}
We present the results of a survey of \aco\ 
emission in 14 infrared luminous dusty star forming galaxies (DSFGs) at $2 < z < 4$ with the NSF's Karl G. Jansky Very Large Array.
All sources are detected in $\rm ^{12}$CO(1$-$0), with an $\sim 1\arcsec$ angular resolution. Seven sources show extended and complex structure. We measure $\rm ^{12}$CO luminosities of $(\mu)L^\prime_{\rm CO(1-0)} = 0.4-2.9\times10^{11}$\,K\,km\,s$^{-1}$\,pc$^2$, and molecular gas masses of $(\mu)M_{\rm H_2} = 1.3-8.6\times10^{11}$\,M$_\odot$, where $(\mu)$ is the magnification factor. The derived molecular gas depletion times of $t_{\rm dep} = 40-460$\,Myr, cover the expected range of both normal star forming galaxies and starbursts. Comparing to the higher$-J$ $\rm ^{12}$CO transitions previously observed for the same sources, we find CO temperature brightness ratios of $r_{32/10} = 0.4-1.4$, $r_{43/10} = 0.4-1.7$, and $r_{54/10} = 0.3-1.3$.  We find a wide range of CO spectral line energy distributions (SLEDs), in agreement with other high-$z$ DSFGs, with the exception of three sources that are most comparable to the Cloverleaf and APM08279$+$5255. Based on radiative transfer modelling of the $\rm ^{12}$CO SLEDs we determine densities of $\rm n_{H_2} = 0.3-8.5\times10^3$\,cm$^{-3}$ and temperatures of $\rm T_K = 100 - 200\,K$. Lastly, four sources are detected in the continuum, three have radio emission consistent with their infrared derived star formation rates, while HerBS-70E requires an additional synchrotron radiation component from an active galactic nucleus.  
 Overall, we find that even though the sample is similarly luminous in the infrared, by tracing the \aco\ emission a diversity of galaxy and excitation properties are revealed, demonstrating the importance of \aco\ observations in combination to higher-$J$ transitions.
 
\end{abstract}

\section{Introduction} \label{sec:intro}

Dusty star-forming galaxies (DSFGs) have played an important role in galaxy growth at high-$z$, as they dominate the cosmic star formation rate density (SFRD) up to redshifts of $z\sim4$ \citep[e.g.,][]{Magnelli2013, Bourne2017, Bouwens2016, Bouwens2020, Hatsukade2018,Zavala2021}, including the peak of the SFRD at $z=1-3$ \citep[see][]{MadauDickinson2014}. The most luminous DSFGs in the infrared (IR) can have luminosities at rest-frame $8-1000\,\mu$m of $L_{\rm IR}>10^{12} - 10^{13}$\,L$_\odot$, corresponding to some of the most intense episodes of star formation activity, with star formation rates (SFRs) that can exceed 1000\,M$_\odot$\,yr$^{-1}$ \citep[see reviews by][]{Blain2002,Casey2014,Hodge2020}.

The number of detected DSFGs at high-$z$ has been steadily increasing over the last decades, with the advent of large area surveys, including the all-sky Planck-HFI \citep{PlanckXXVII2015, Canameras2015}, and the South Pole Telescope \citep[SPT;][]{Carlstrom2011} surveys \citep{Vieira2010,Vieira2013}, amongst others.
The large field surveys on the Herschel Space Observatory \citep[{\em Herschel};][]{Pilbratt2010} covering $\rm >1000 \, deg^2$ of the extragalactic sky at $\rm 70-500 \,\mu m$, have significantly contributed to the increased number of known DSFGs. Surveys such as 
the {\em Herschel} Astrophysical Terahertz Large Area Survey \citep[H-ATLAS;][]{Eales2010}, the {\em Herschel} Multi-tiered Extragalactic Survey \citep[HerMES;][]{Oliver2012}, and the {\em Herschel} Stripe 82 Survey \citep[HerS;][]{Viero2014},
have led to the detection of $> 10^5$ DSFGs, including $>200$ luminous DSFGs with $S_{\rm 500 \, \mu m} = 80-900 \, \rm mJy$ \citep[e.g.,][and references therein]{Nayyeri2016,Bakx2018}.

Dedicated follow-up studies of the luminous DSFGs identified with {\it Herschel}, have found that these sources cover a wide redshift range of $1<z<6$ \citep[e.g.,][and references therein]{Nayyeri2016,Bakx2018}, include numerous gravitationally amplified galaxies \citep[e.g.,][]{Negrello2010, Negrello2017, Conley2011, Riechers2011d, Cox2011, Wardlow2013, Bussmann2013, Nayyeri2016,Bakx2020a,Bakx2020b}, rare galaxies with $L_{\rm IR} > 10^{13} \, \Lsun$ classified as hyper-luminous infrared galaxies (HyLIRGs) \citep[e.g., ][]{Ivison2013, Ivison2019, Fu2013, Oteo2016, Riechers2013, Riechers2017}, 
and over-densities of DSFGs that are blended due to the large {\em Herschel} beam 
\citep[e.g.,][]{Bussmann2015, Oteo2018, Gomez-Guijarro2019, Ivison2019}. 
The diversity and size of the {\it Herschel} selected DSFG population makes it ideal for dedicated follow-up surveys of large statistical samples 
to determine the galaxy properties and nature of bright DSFGs at high-$z$. 

Tracing the molecular gas in the interstellar medium (ISM) of these galaxies is of particular interest, as it is the fuel for star formation. Direct observations of the
$\rm ^{12}$CO(1$-$0) transition are necessary to measure total molecular gas masses, and depletion timescales, probe the cold gas distribution and morphology, and anchor the modeling of the $\rm ^{12}$CO spectral line energy distributions (SLEDs) from which the physical properties of the cold gas (e.g., kinetic temperature, and density) can be derived \citep[e.g.,][]{Ivison2011,Riechers2011e,Riechers2011d,Riechers2011c,Harris2012,Sharon2016,Aravena2016}. 

A first necessary step towards this direction is the robust determination of the redshifts of these sources. Ongoing efforts have lead to the measurement of spectroscopic redshifts for over $\sim300$ of the brightest high-$z$ DSFGs 
\citep[e.g.,][]{Weiss2009,Walter2012,Harris2012,Lupu2012,Weiss2013,Strandet2016,Fudamoto2017,Danielson2017,Reuter2020,Urquhart2022}, with $z$-GAL \citep[][Cox et al. in prep]{Neri2020} being the largest redshift survey among them.

%zGAL pilot project
The first series of $z$-GAL sources for which reliable spectroscopic redshifts were determined, was presented in \cite{Neri2020}. In total 14 individual high-$z$ luminous DSFGs were detected with NOEMA in 11 target fields that were {\it Herschel} selected based on their 500$\mu$m fluxes. The sources were detected either in multiple $\rm ^{12}$CO lines or in a combination of $\rm ^{12}$CO and other molecular or atomic species. However, despite the multi-line detections, the conclusions of this study on the molecular gas properties of the sample remained limited due to the lack of information on the lowest-$J$ $\rm ^{12}$CO lines. This motivated targeted $\rm ^{12}$CO(1$-$0) follow-up of these galaxies with the NSF's Karl G. Jansky Very Large Array (VLA).

%This paper
In this paper, we present the survey of $\rm ^{12}$\aco\ emission and the underlying radio continuum of 14 luminous DSFGs from the $z$-GAL Pilot Program in the redshift range of  $2<z<4$. In Section~\ref{sec:obs}, we describe the sample, the observations and the data reduction. In Section~\ref{sec:results}, we report the main results from the VLA observations both for the $\rm ^{12}$CO(1$-$0) emission line and the continuum. Section~\ref{sec:discussion} outlines the implications of these results; particularly, addressing the total gas molecular gas mass, the excitation conditions and the nature of these sources. Finally, Section~\ref{sec:conclusions} summarizes the main findings of this paper. Throughout this paper, we adopt a spatially flat $\Lambda$CDM cosmology with $H_{0}=67.4\,{\rm km\,s^{-1}\,Mpc^{-1}}$ and $\Omega_\mathrm{M}=0.315$ \citep{Planck2020}.

\begin{table*}[th]
\begin{center}
\caption{The $z$-GAL Pilot Program sources}
\begin{tabular}{lcccccc}
\hline
\hline
Source & RA & Dec. &  $z_{\rm spec}$ & $\rm \Delta V$ & $\mu_L L_{\rm IR}$ &  $\mu_L M_{\rm dust}$ \\ 
        & \multicolumn{2}{c}{(J2000)} & & $\rm (km \, s^{-1})$ &  $10^{12} \, L_\odot$ & $10^{10} \, M_\odot$ \\ 
\hline
HerBS-34   & 13:34:13.9 & 26:04:57.5 &  2.663 &  330$\pm$10  & 40.6$\pm$2.4 & 1.16$\pm$0.07 \\
HerBS-43a  & 13:24:18.8 & 32:07:54.4 &  3.212 & 1070$\pm$90  & 33.4$\pm$4.1 & 0.69$\pm$0.04 \\ 
HerBS-43b  & 13:24:19.2 & 32:07:49.2 &  4.054 &  800$\pm$50  & 15.0$\pm$2.0 & 0.45$\pm$0.04 \\
HerBS-44   & 13:32:55.8 & 34:22:08.4 &  2.927 &  520$\pm$50  & 108.1$\pm$7.8 & 0.7$\pm$0.04 \\
HerBS-54   & 13:15:40.7 & 26:23:19.6 &  2.442 & 1020$\pm$190 & 23.6$\pm$1.6 & 1.25$\pm$0.12 \\ 
HerBS-58   & 13:03:33.2 & 24:46:42.3 &  2.084 &  970$\pm$50  & 17.7$\pm$1.3 & 0.83$\pm$0.06 \\ 
HerBS-70E  & 13:01:40.3 & 29:29:16.2 &  2.308 &  770$\pm$50  & 40.5$\pm$5.2 & 0.37$\pm$0.02 \\
HerBS-70W  & 13:01:39.3 & 29:29:25.2 &  2.311 &  140$\pm$20  & 8.7$\pm$4.2 & 0.09$\pm$0.02 \\ 
HerBS-79   & 13:14:34.1 & 33:52:20.1 &  2.078 &  870$\pm$70  & 18.1$\pm$1.2 & 0.83$\pm$0.18 \\
HerBS-89a  & 13:16:11.5 & 28:12:17.7 &  2.949 & 1080$\pm$60  & 28.9$\pm$2.5 & 1.30$\pm$0.07 \\ 
HerBS-95E  & 13:43:42.7 & 26:39:18.0 &  2.972 &  870$\pm$50  & 11.4$\pm$1.9 & 0.50$\pm$0.05 \\
HerBS-95W  & 13:43:41.5 & 26:39:22.7 &  2.973&   540$\pm$30  & 16.1$\pm$1.8 & 0.77$\pm$0.04 \\
HerBS-113  & 13:12:11.3 & 32:38:37.8 &  2.787 &  900$\pm$200 & 29.6$\pm$2.8 & 0.72$\pm$0.06 \\
HerBS-154  & 13:22:58.1 & 32:50:51.7 &  3.707 &  310$\pm$40  & 81.3$\pm$7.3 & 0.46$\pm$0.03 \\ 
\hline
\end{tabular}
\tablenotetext{}{{\bf Notes.} The coordinates and properties of the sources are taken from \citet{Neri2020}, with the exception 
of HerBS-89a for which we include the revised values reported in \citet{Berta2021}. None of the properties in this table have been corrected for gravitational magnification ($\mu_L$ is the magnification factor, assuming no differential lensing between the CO and dust emission). The $\rm \Delta V$ corresponds to the mean FWHM of the CO transitions observed. The $\rm 8-1000 \, \mu m$ rest-frame infrared luminosities ($L_{\rm IR}$) and dust masses ($M_{\rm dust}$) are those derived using the \citet{DraineLi2007} approach - see \citet{Neri2020} and \citet{Berta2021} for further details.}
\label{tab:Sample}
	
\end{center}
\end{table*}

\section{Sample, Observations \& Analysis} \label{sec:obs}

\subsection{Sample}
The sources of the Pilot Program \citep{Neri2020} were selected from the {\it Herschel} Bright Sources (HerBS) sample \cite[]{Bakx2018}; they are located close to the North Galactic pole (NGP) and have flux densities in the range $\rm 80\, mJy  \lesssim S_{500 \, \mu m} \lesssim 130 \, mJy$.  From the 13 {\it Herschel} targets observed with NOEMA for the Pilot Program, three were resolved into multiple sources. HerBS-70 and HerBS-95 were resolved into binary systems, and HerBS-43 was resolved into two galaxies at different redshifts. This resulted in a total of 16 individual galaxies that were presented in \citet{Neri2020}, 14 of which have a reliable redshift in the range of $2.08 < z <4.05$ and were selected for follow-up VLA observations. These sources are spatially unresolved or barely resolved with the current NOEMA data with angular resolutions between $1\farcs2$ and $6\arcsec$ \citep{Neri2020}.

\begin{table*}[th]
\begin{center}
\caption{Observed properties of the \aco\ data} \label{tab:sources}
\begin{tabular}{lccccccccc}
\hline 
\hline 
Source & \multicolumn{2}{c}{Moment-0 map properties } & & \multicolumn{5}{c}{\aco\ line properties} \\ \cline{2-3}  \cline{5-9}

       & rms$^a$       & beam$^b$ &  & spec-rms$^c$ & $S_{\rm peak}^d$ & $\rm FWHM^e$ & $I_{\rm CO(1-0)}^f$ & Extent of emission$^g$  \\ 
       & [Jy\,km\,s$^{-1}$\,beam$^{-1}$]     &   [arcsec$^2$] &  &  [mJy] &  [mJy]      & [km\,s$^{-1}$] & [Jy\,km\,s$^{-1}$] & [arcsec$^2$] \\ 
\hline 
HerBS-34  & 0.055  & $0.8\times0.7$  & & $0.19$ & $0.7\pm0.12$ & $593\pm112$ & $0.44\pm0.11$ & $(1.7\pm0.4)\times(0.9\pm0.2)$\\ 
HerBS-43a & 0.048  & $1.0\times0.9$  & & $0.13$ &$0.3\pm0.06$  & $1166\pm249$ & $0.37\pm0.10$ & $(1.2\pm0.2)\times(0.9\pm0.3)$\\ 
HerBS-43b & 0.016  & $1.3\times1.0$  & & $0.05$ &$0.08\pm0.03$ & $744\pm290$  & $0.064\pm0.033$ & $(1.1\pm0.3)\times(0.3\pm0.6)$  \\ 
HerBS-44  & 0.047  & $0.9\times0.8$   & & $0.18$ &$0.95\pm0.14$ & $377\pm63$   & $0.38\pm0.08$ & $(1.1\pm0.3)\times(0.5\pm0.2)$\\ 
HerBS-54  & 0.055  & $0.8\times0.7$  & & $0.25$ &$0.8\pm0.11$  & $1087\pm176$ & $0.92\pm0.19$ & $\lesssim2.2\times2.1$\\ 
HerBS-58  & 0.046  & $0.9\times0.6$  & & $0.57$ &$2.12\pm0.32$ & $363\pm64$   & $0.82\pm0.19$ & $(2.1\pm0.4)\times(1.6\pm0.3)$\\ 
HerBS-70e & 0.03   & $0.9\times0.7$  & & $0.15$ &$0.49\pm0.09$ & $622\pm130$  & $0.32\pm0.09$ & $(1.6\pm0.3)\times(1.2\pm0.3)$ \\ 
HerBS-70w & 0.019  & $0.9\times0.7$  & & $0.15$ &$0.83\pm0.15$ & $197\pm39$   & $0.17\pm0.04$ & $(2.1\pm0.5)\times(1.1\pm0.3)$ \\ 
HerBS-79  & 0.043  & $0.8\times0.6$  & & $0.34$ &$1.24\pm0.17$ & $787\pm125$  & $1.04\pm0.22$ & $(2.9\pm0.5)\times(1.0\pm0.2)$\\ 
HerBS-89a & 0.07  &  1.2 $\times$ 0.8  & & $0.24$ & $0.64\pm0.09$ & $1586\pm247$ & $1.08\pm0.22$ & $(2.1\pm0.5)\times(1.3\pm0.4)$\\
HerBS-95e & 0.02   & $1.0\times0.8$  & & $0.06$ & $0.2\pm0.04$  & $658\pm137$  & $0.14\pm0.04$ & $(1.5\pm0.3)\times(0.8\pm0.2)$ \\ 
HerBS-95w & 0.024  & $1.0\times0.8$  & & $0.19$ & $0.95\pm0.12$ & $522\pm76$   & $0.52\pm0.10$ & $(2.5\pm0.4)\times(1.9\pm0.3)$\\ 
HerBS-113 & 0.035  & $1.0\times0.7$  & & $0.32$ & $1.46\pm0.21$ & $497\pm81$   & $0.77\pm0.16$ & $\lesssim3\times2.8$\\ 
HerBS-154 & 0.035  & $1.2\times1.0$  & & $0.16$ & $0.95\pm0.11$ & $384\pm54$   & $0.39\pm0.07$ & $(2.7\pm0.4)\times(1.7\pm0.3)$ \\ 
\hline
\end{tabular}
\tablenotetext{}{{\bf Notes.} (a) The rms of the moment-0 maps; (b) the beam sizes of the moment-0 maps; (c) the rms per 100\,km/s channels of the extracted spectra, based on the line free channels; (d) the fitted flux peak of the \aco\ emission line; (e) the fitted full width half-max (FWHM) of the \aco\ emission line from single Gaussian fits; (f) the integrated line flux derived from the fit to the line; (g) estimates using {\sc imfit} on the size of the \aco\ emission are based on the moment-0 maps, deconvolved from the beam, with the exception of HerBS-54 and HerBS-113 for which we were not able to use the single gaussian fit of {\sc imfit} due to their complexity. For HerBS-54 and HerBS-113 we instead provide a rough upper limit of the extent of the emission.
}
\end{center}
\end{table*}

\subsection{VLA observations \& data reduction}

We observed the $\rm ^{12}$CO(1$-$0) - hereafter \aco\ - emission line for the 14 luminous DSFGs with reliable redshifts from the $z$-GAL Pilot Program (Table~\ref{tab:Sample}). Observations were carried out using the NSF's Karl G. Jamsky Very Large Array (VLA), 
in the C configuration (program I.D.:VLA/20A-083 - P.I.: D. Riechers). We used the 0.9~cm Ka and K band, with the correlator configured to an 8-bit sampling mode, achieving a spectral resolution of 2\,MHz (i.e., $\rm 16-21 \, km\, s^{-1}$), over a total bandwidth of 2\,GHz in dual polarization.
The observations were set up so that one of the two side-bands was centred on the expected frequency of the redshifted \aco\ emission line ($\nu_{\rm rest}=115.271 \, \rm GHz$) for each source. The second sideband was either positioned alongside the first to provide contiguous frequency coverage and maximise continuum sensitivity or re-positioned to a significantly different frequency to obtain data with which to measure the continuum spectral index and cover faint spectral lines. Three target fields were observed with the former setting, and nine with the latter. However, the integration times were chosen to detect the \aco\ emission, not the underlying continuum. Note that HerBS-43a was also in the field of view for the observations of HerBS-43b, and was observed with two different frequency tunings. 
The data were acquired in January-May 2020 under stable atmospheric conditions and each source was observed for between $1-7.6$\,hours with $0.6-5.4$\,hours on-source integration. 
The 2~GHz bandwidth setup was used to maximize the potential for stacking of faint lines, while at the same time retaining sufficient spectral resolution (2~MHz) to finely sample the CO emission line for each source. The 8-bit samplers were selected to maximize sensitivity. The gain, bandpass, and flux calibrators used were 3C286, J1327$+$2210, J1310$+$3220, and J1310$+$3230. 

The data were reduced, calibrated and imaged using {\sc CASA v5.6.2} (Common Astronomy Software Application\footnote{https://casa.nrao.edu}; \citealt{McMullin2007}). Manual data flagging during reduction was necessary for most sources. 
The accuracy of the flux density calibration is within 10--15\%, when comparing the measured fluxes for the calibrators from our observations to the calibrator models, with a median deviation of 3.33\%.

Spectral cubes, continuum, and \aco\ moment-0 maps were created and cleaned using the {\sc tclean} function of {\sc casa} with a natural weighting. The spectral channel width in the cubes was set to 100\,km\,s$^{-1}$ and continuum maps were created using the line-free channels. The resulting moment-0 maps have spatial resolutions varying from $0\farcs9 \times 0\farcs6$ to $1\farcs2 \times 1\farcs0$ and rms noise levels between 16 and $\rm 55 \, \mu Jy\,km\,s^{-1}$\,beam$^{-1}$ (Table~\ref{tab:sources}). For those sources where the underlying continuum is detected, we performed continuum subtraction for the spectral cubes using the line-free channels. Two continuum images were produced for the sources, with the exception of the three sources with continuous frequency coverage between the two sidebands for which only a single continuum image is produced.

\subsection{Spectral analysis}

In order to obtain the best quality \aco\ spectra, we follow an iterative process for defining the extraction region used. Starting with a $1\arcsec$ circular aperture centered on the source position, we extract initial spectra, from which we define the line channels to be used for the moment-0 maps. Once the moment-0 map is created, we define new extraction regions based on the 2$\sigma$ contours of the map. We repeat the steps of selecting the line channels to be used for the moment-0 maps and defining the 2$\sigma$ contour extraction region, until the 2$\sigma$ contour converges. To achieve a higher SNR and retrieve the full information on the extent of the emission as best as possible, the total line intensity maps were created directly from the uv data and cleaned using {\sc tclean} in {\sc casa}. 
In Figure~\ref{fig:spec}, we present the extracted spectra and moment-0 maps for each source.  
The properties of the extracted \aco\ lines are listed in Table~\ref{tab:sources} together with the rms noise levels and the spatial resolution of 
the moment-0 maps. For sources where a clear double peak is seen in the line profile, we fit with both single and double Gaussian components; however, we find that the integrated flux does not change significantly between the two. As the more complex fit is not required by the data and will only increase the uncertainties due to the larger number of fitting parameters, we chose to only use the results of the simpler, single Gaussian fit.

\subsection{Continuum analysis}
\begin{figure*}
\centering
\includegraphics[width=\textwidth]{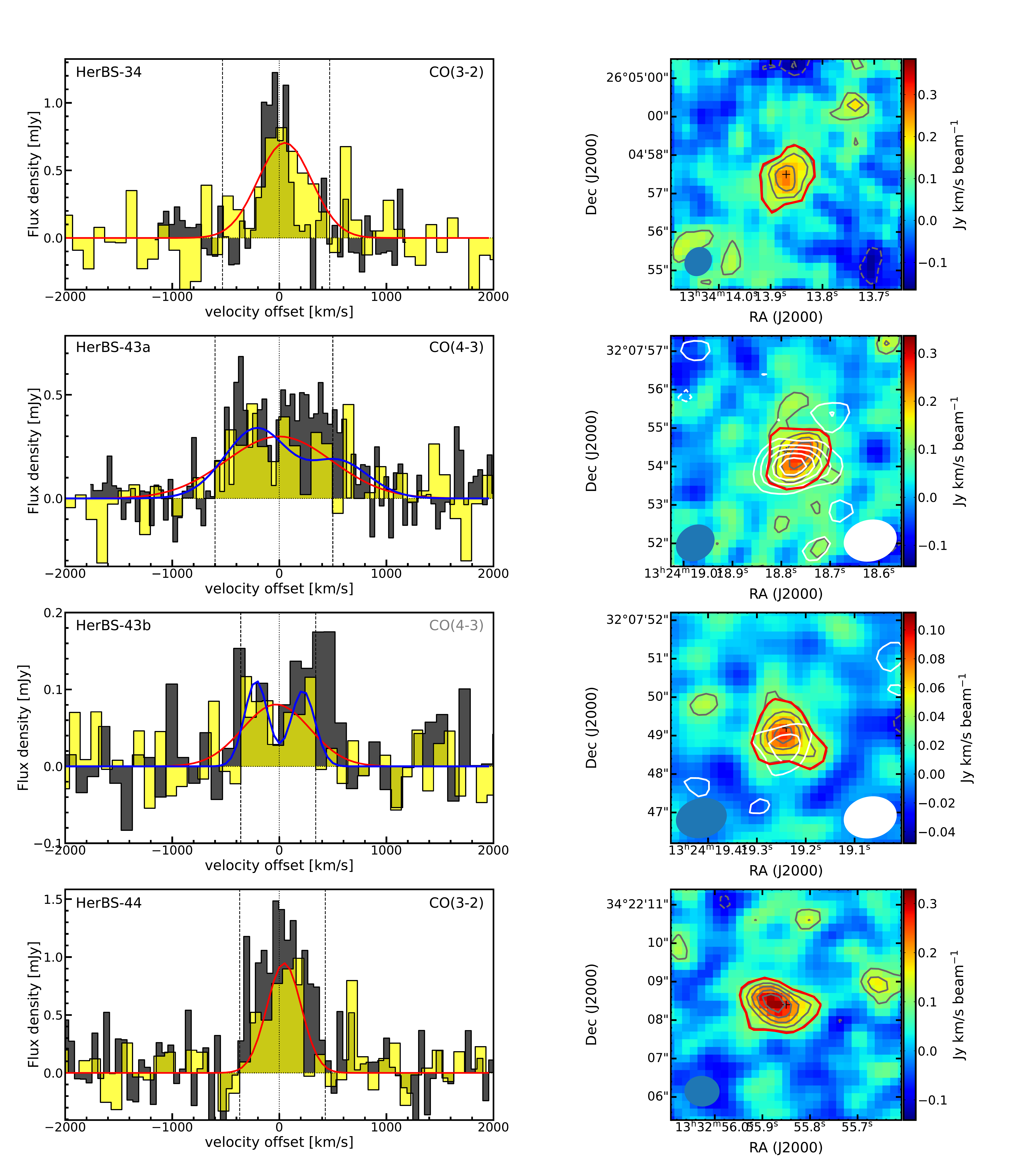}
\caption{\aco\ spectra (left), moment-0 (right) for each source in our sample. The source names are indicated on the top of the left panels. Each \aco\ emission line is plotted in yellow and centered at the zero velocity corresponding to the redshift of the source (see Table~\ref{tab:Sample}), and the Gaussian fits to the lines are shown with red curves. For comparison, we plot the next lowest-$J$ CO line from \cite{Neri2020} in grey, normalized to 1.5$\times$ the peak of the \aco\ line to show the respective line profiles with more clarity. The comparison line used is indicated on the top right of each plot. The moment-0 map contours are plotted starting at $2\sigma$ and increase in steps of $1\sigma$, where the $1\sigma$ noise levels for each source are listed in Table~\ref{tab:sources}. The symmetric negative contours are also shown, with dashed curves. The synthesized beam is shown in the lower left corner of each moment-0 map. The maps are centered on the positions listed in Table~\ref{tab:Sample}. For the sources detected in continuum (HerBS-43a, 43b, 54, and 70E), we overplot in white contours the continuum emission on the moment-0 maps, and the corresponding synthesized beam on the lower right corner. The continuum contours start at $2\sigma$ and increase in steps of $1\sigma$, with the exception of HerBS-70E for which we plot contours at 2, 5, 10, 30, and 40$\sigma$ (see Table~\ref{tab:cont}). The red polygon on the moment-0 maps corresponds to the region from which we extracted the spectra.} \label{fig:spec}. 
\end{figure*}

As the aim of the observations was to detect the \aco\ emission, and the observations were not designed to detect the continuum, the majority of our sample is not detected in the continuum, with the exception of four sources, namely HerBS-43a, HerBS-43b, HerBS-54 and HerBS-70E. For these sources, we used the 2D Gaussian fitting tool {\sc imfit} of {\sc casa} to retrieve the total continuum flux density. In Table~\ref{tab:cont}, we give the continuum flux densities, the rms values, and the corresponding frequencies. The radio continuum emission for the four sources detected are displayed as white contours on the moment-0 maps of Fig.~\ref{fig:spec}. For the sources where no continuum emission was detected, Table~\ref{tab:cont} lists the rms noise level of each map. 

Within the field of HerBS-44, a serendipitous source is detected in the radio continuum (at 29 and 38~GHz) at a distance of $\sim34\arcsec$ from the main source. The position of the source is at RA: 13:32:57.8, Dec: +34:22:27.0 (J2000.0), and the continuum fluxes are reported in Table~\ref{tab:cont} under the source name HerBS-44s. A likely counterpart to this source is FIRST J133257.7+342227 with a separation of 0.36$\arcsec$, or NVSS J133257+342228 at a separation of 1$\arcsec$.

\begin{figure*}
\centering
\includegraphics[width=\textwidth]{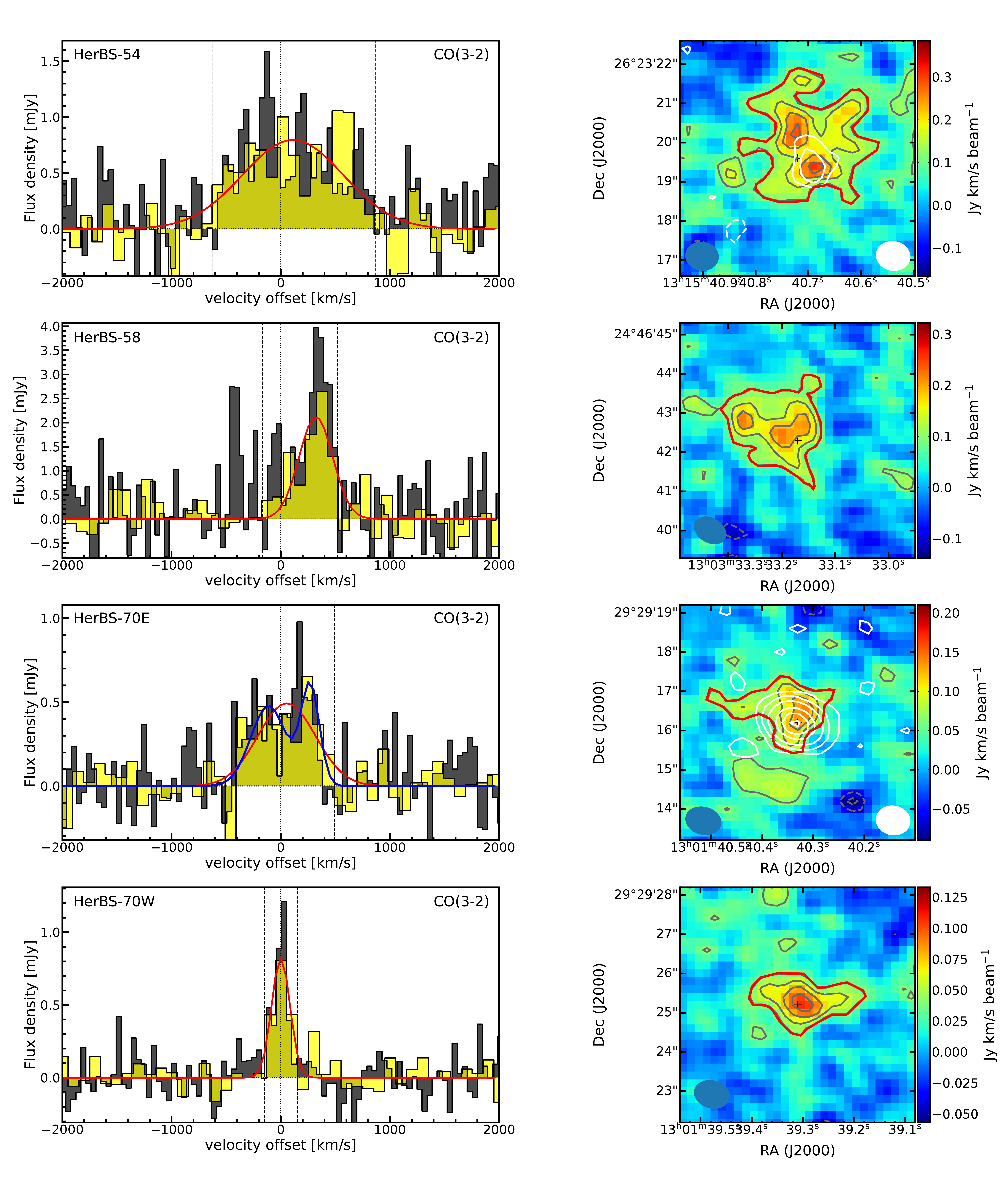}
{\bf Figure 1 (continued)}
\end{figure*}

\begin{figure*}
\centering
\includegraphics[width=\textwidth]{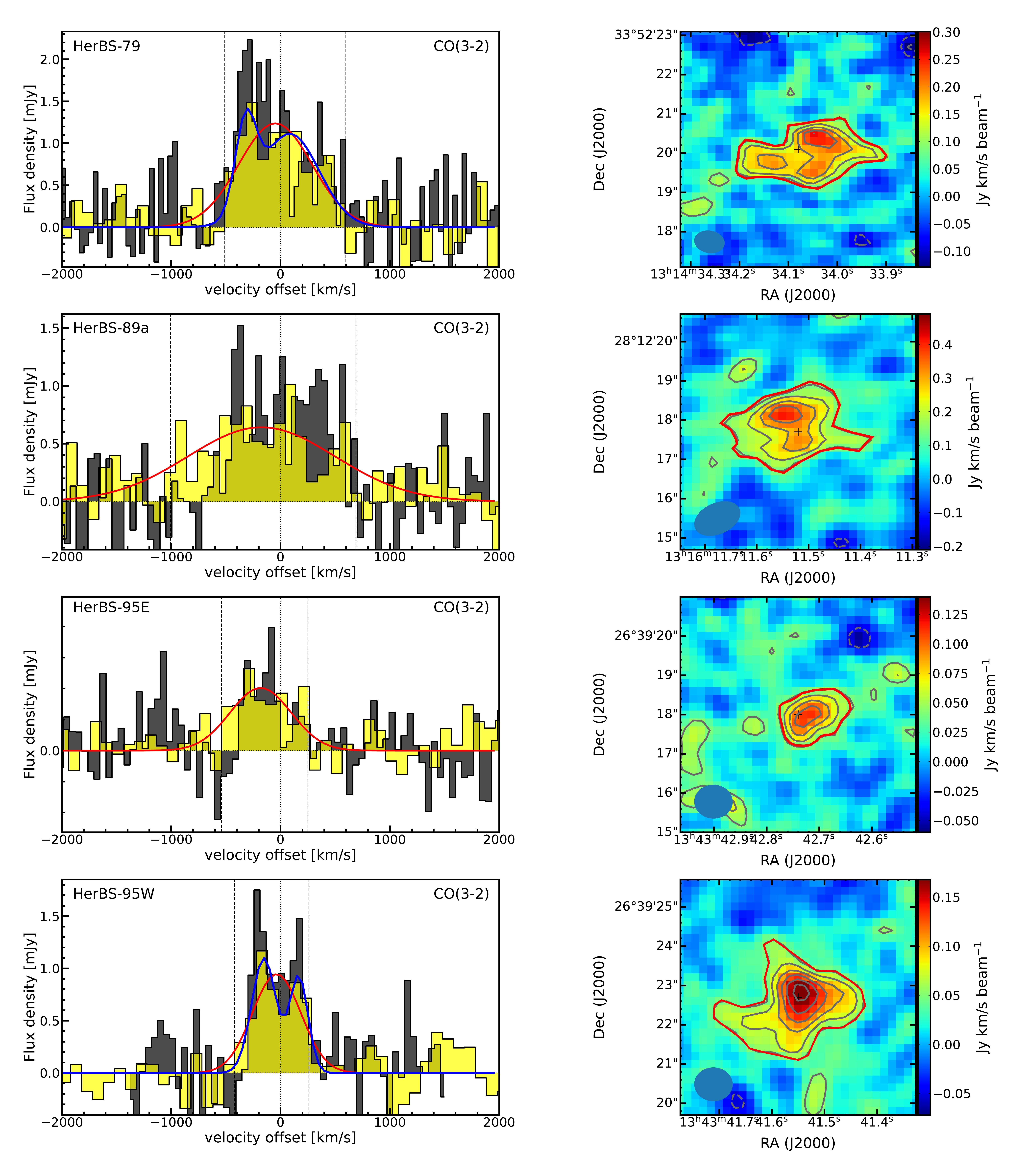}
{\bf Figure 1 (continued)}
\end{figure*}

\begin{figure*}
\centering
\includegraphics[width=\textwidth]{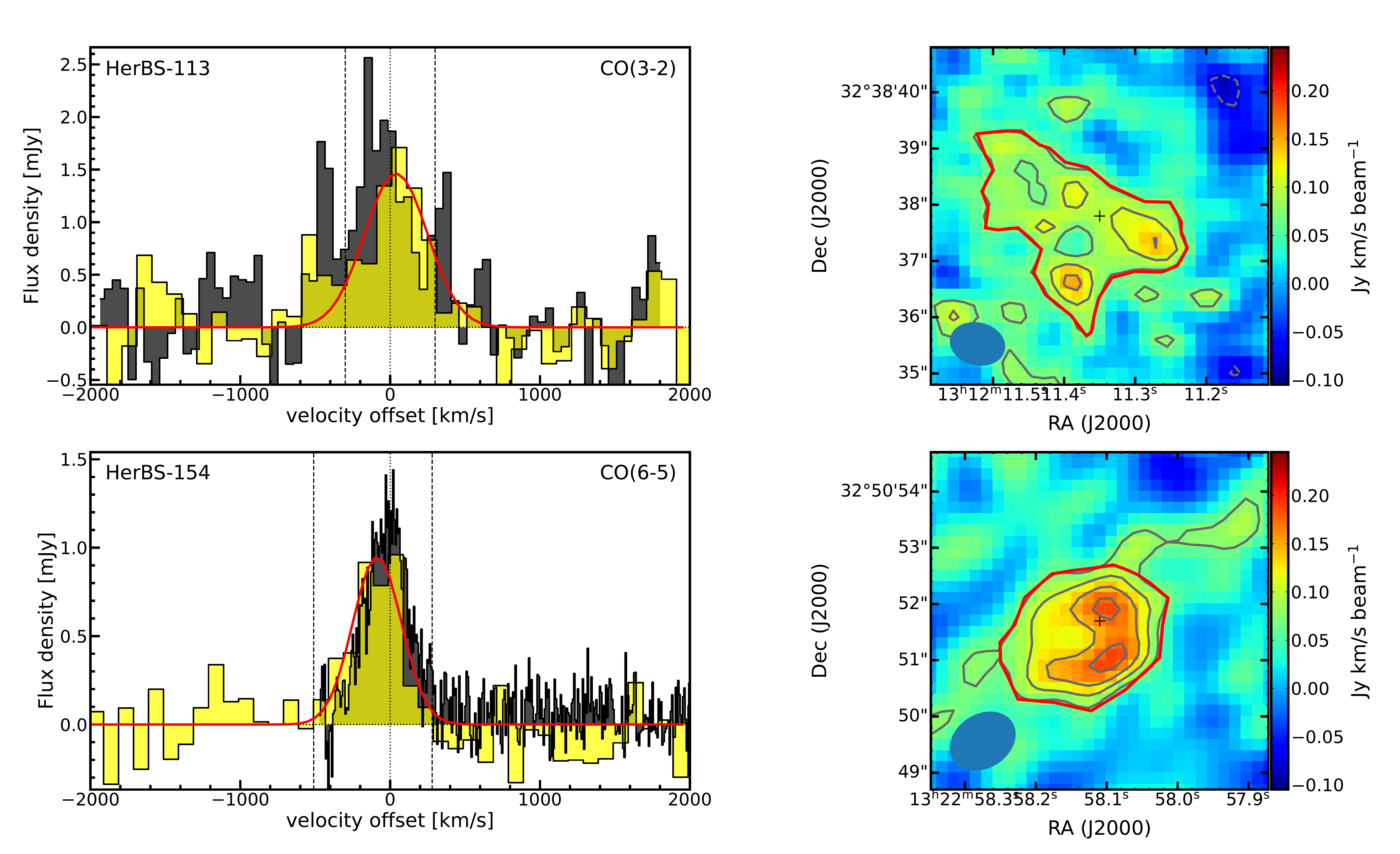}
{\bf Figure 1 (continued)}
\end{figure*}

\begin{figure}
\centering
\includegraphics[width=0.45\textwidth]{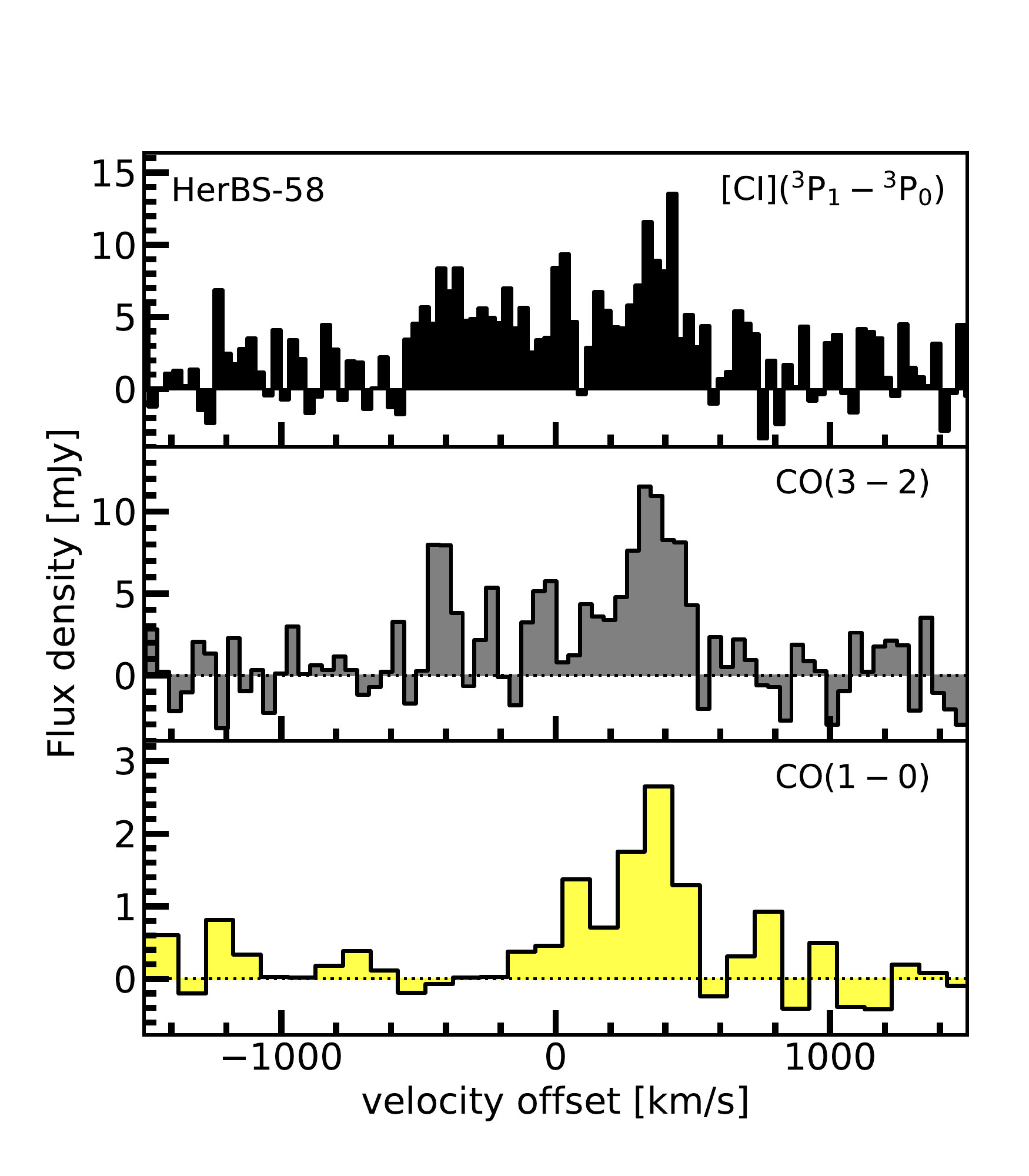}
\caption{A comparison of the detected emission lines of HerBS-58. The lines plotted are indicated on the upper left of each panel. Both the \cco\ and [CI] spectra have additional emission at negative velocities, not seen in the \aco\ spectra.}\label{fig:herbs58}
\end{figure}

\begin{figure*}
\centering
\includegraphics[width=0.45\textwidth]{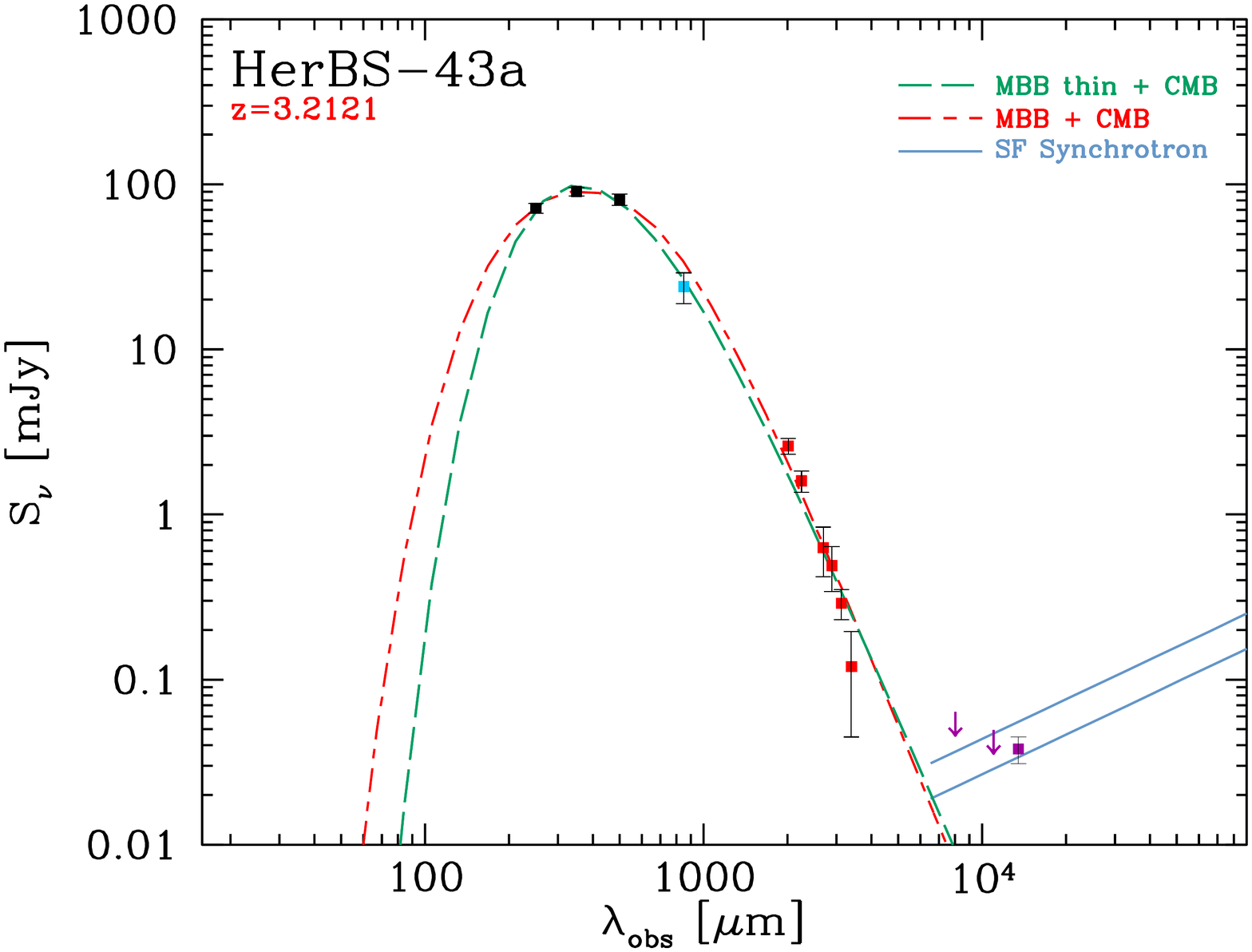}
\includegraphics[width=0.45\textwidth]{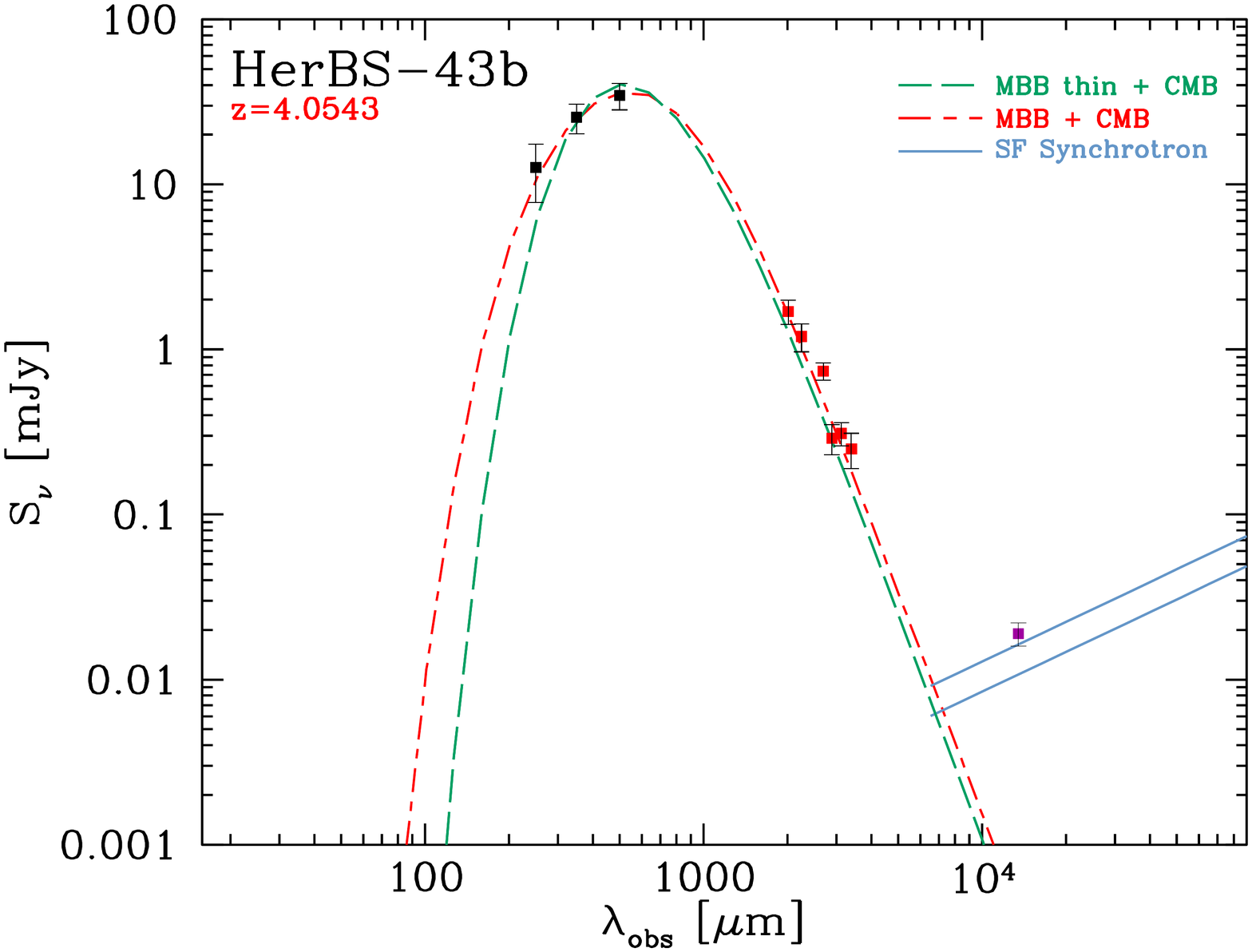}\\
\includegraphics[width=0.45\textwidth]{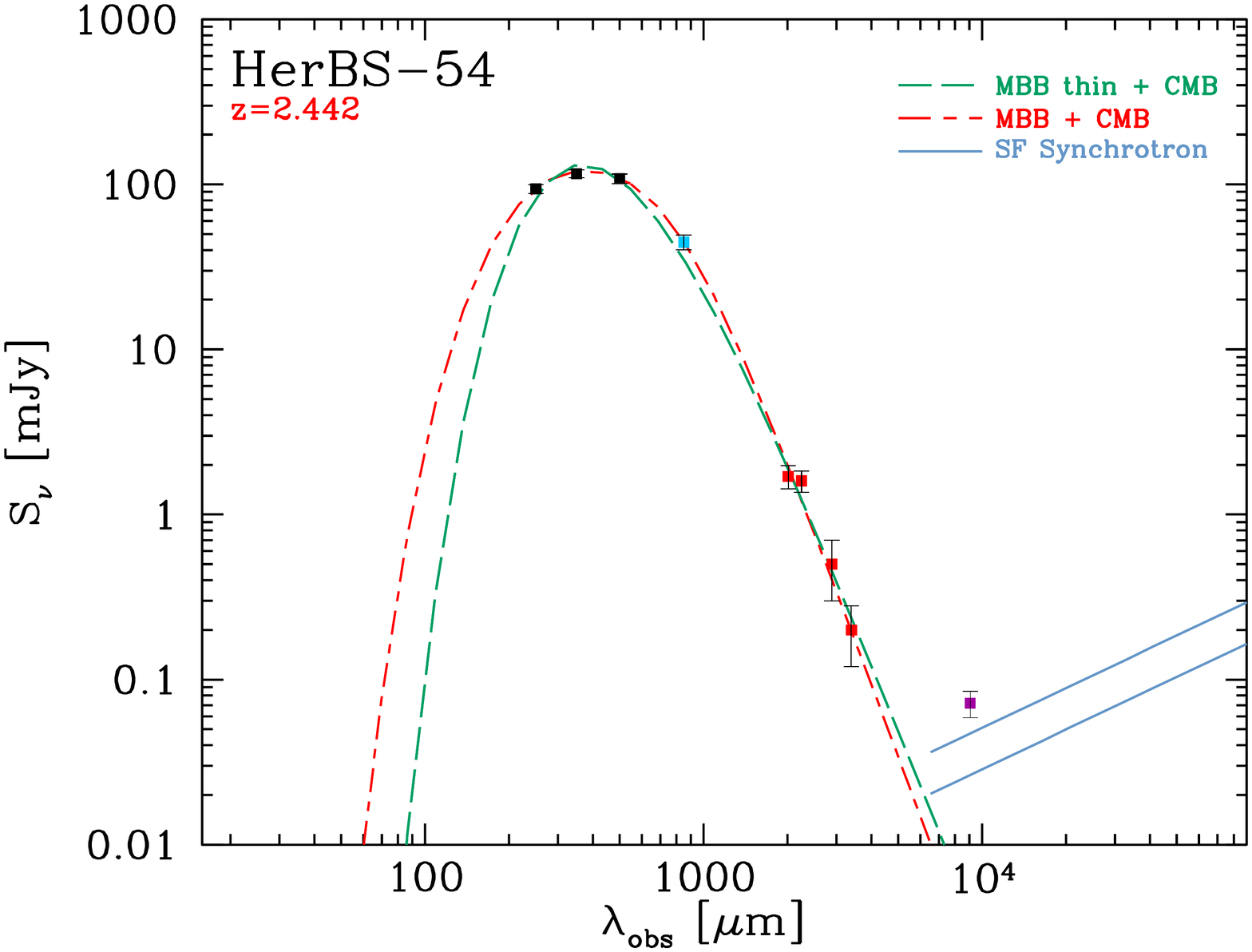}
\includegraphics[width=0.45\textwidth]{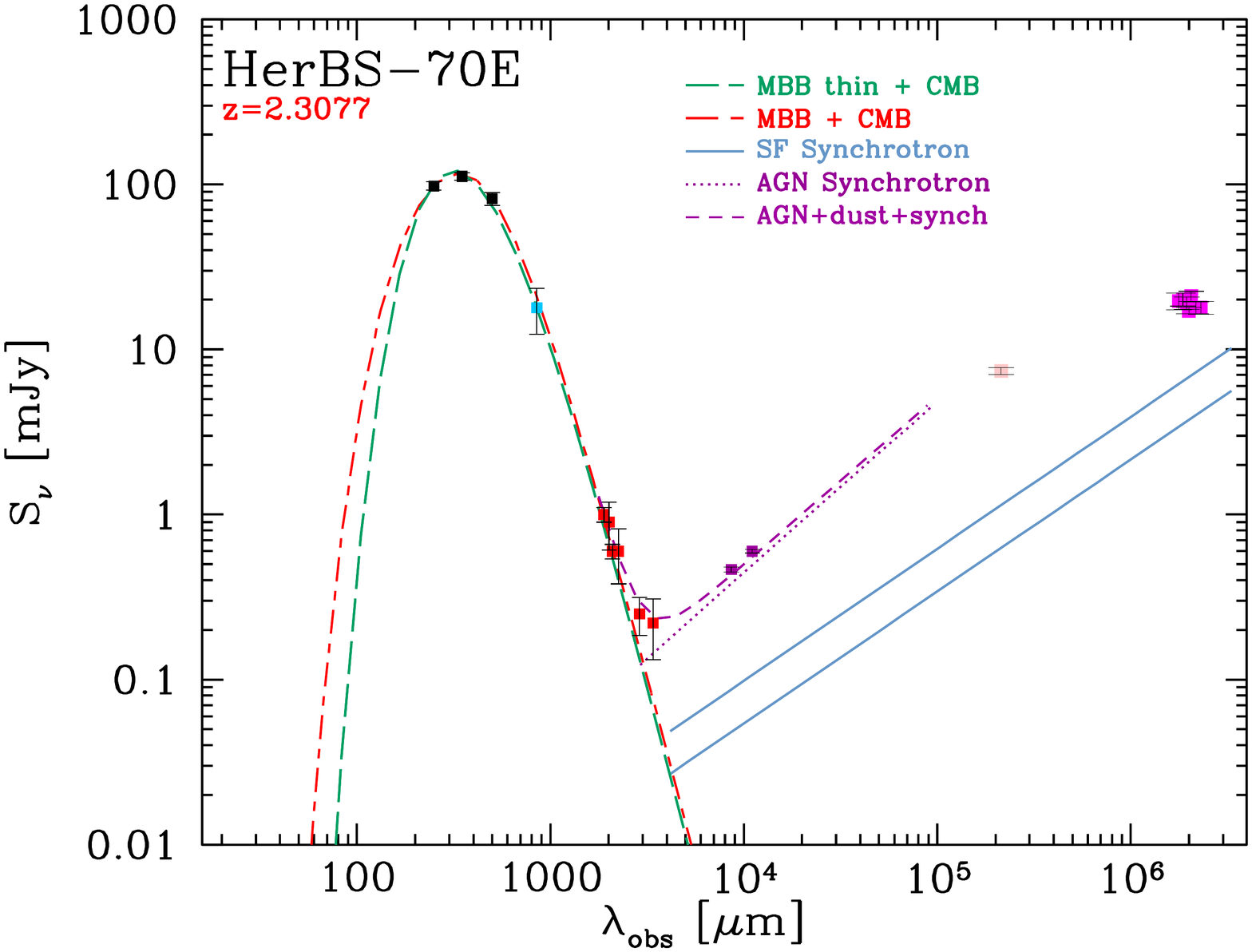}
\caption{Observed spectral energy distributions for the four sources detected in the radio continuum, HerBS-43a, HerBS-43b, HerBS-54, and HerBS-70E (from top left to bottom right panel). 
The data include SPIRE and SCUBA-2 flux densities \citep[black and blue dots;][]{Bakx2018} and NOEMA continuum flux densities at 2 and 3~mm \citep[red dots;][]{Neri2020}; the VLA measurements at 22--38\,GHz (13--7.8\,mm) are shown as purple dots and the 3~$\sigma$ upper limits as purple arrows. The fits are based on modified black body (MBB) dust models [green and red lines] including corrections for the effects of the CMB (see text for details). The light-blue continuous lines represent the synchrotron emission due to star formation, with a fixed spectral index $\alpha=-0.8$, normalised on the basis of two radio/far-infrared correlations derived by \citet{Delhaize2017} (lower line), and \citet{Magnelli2015} (upper line). In the case of HerBS-70E, we also plot the synchrotron emission due to the AGN (purple dotted line), and the sum of all components (purple dashed line), and additional radio data from the FIRST survey at 1.4\,GHz and LOFAR at 126-173~MHz, for comparison.
} 
\label{fig:radio-continuum}
\end{figure*}

\begin{table*}[th]
\begin{center}
\caption{Continuum measurements}\label{tab:cont} 
\begin{tabular}{lcccccc}
\hline 
\hline 
Source  & \multicolumn{2}{c}{$\nu_{\rm cen}$}  & \multicolumn{2}{c}{rms} & \multicolumn{2}{c}{S$_{\rm cont}$} \\
&\multicolumn{2}{c}{[GHz]} & \multicolumn{2}{c}{[mJy/beam]} & \multicolumn{2}{c}{[mJy]} \\
 & LSB & USB & LSB & USB & LSB & USB \\
\hline 
HerBS-34  & \multicolumn{2}{c}{32.02} & \multicolumn{2}{c}{0.011} & \multicolumn{2}{c}{--}  \\ 
HerBS-43a & 27.3 & 37.45 & 0.014 & 0.018 & --  & --   \\ 
HerBS-43a & \multicolumn{2}{c}{22.24} & \multicolumn{2}{c}{0.004}  & \multicolumn{2}{c}{0.024$\pm$0.007} \\
HerBS-43b & \multicolumn{2}{c}{22.24} & \multicolumn{2}{c}{0.004}  & \multicolumn{2}{c}{0.019$\pm$0.003}  \\ 
HerBS-44  & 29.16 & 38.51 & 0.016 & 0.023 & --  & -- \\ 
HerBS-44s & 29.16 & 38.51 & 0.037 & 0.045 & 0.77$\pm$0.04  & 0.65$\pm$0.08 \\
HerBS-54  & \multicolumn{2}{c}{33.06} & \multicolumn{2}{c}{0.009} & \multicolumn{2}{c}{0.072$\pm$0.013}  \\ 
HerBS-58  & 28.80  & 37.31  &  0.014 & 0.02 & -- & --  \\ 
HerBS-70E & 27.11 & 34.78 & 0.01  & 0.01 & 0.60$\pm$0.02 & 0.47$\pm$0.02 \\ 
HerBS-70W & 27.11 & 34.78 & 0.01 & 0.01 & --   & -- \\ 
HerBS-79  & 28.73 & 37.51 & 0.01  & 0.015  & --  & -- \\ 
HerBS-89a & 29.18 & 38.49 & 0.02  & 0.05   & -- & -- \\
HerBS-95E & 28.95 & 38.5 & 0.008 & 0.01 & --  & -- \\ 
HerBS-95W & 28.95 & 38.5 & 0.008 & 0.01  & -- & --   \\ 
HerBS-113 & 30.37 & 38.5 & 0.016 & 0.022 & --  & --  \\ 
HerBS-154 & 19.15 & 24.42 & 0.022 & 0.01 & --  & -- \\ 
\hline
\end{tabular}
\tablenotetext{}{{\bf Notes.} The source HerBS-43a was observed with two different tunings. The results for HerBS-89a are from \citet{Berta2021}. HerBS-44s is a serendipitous source detected in the radio continuum located $\sim34\arcsec$ to the north-east of HerBS-44 at RA: 13:32:57.8, Dec: +34:22:27.0 (J2000.0).}
\end{center}
\end{table*}

\section{Results}\label{sec:results}
In this section we present the initial results on the observed \aco\ line emission (Section~\ref{sec:CO_results}), and the radio continuum measurements (Section~\ref{sec:continuum_results}) of our sample. 

\subsection{Observed emission line properties} \label{sec:CO_results}

All sources are detected in \aco\ with at least $\rm 5 \sigma$ significance for the integrated line flux. Here we provide an overview of the observed properties of our sample, with 
more detailed descriptions for the individual sources reported in Appendix~\ref{app:ind_sources}.

In Fig.~\ref{fig:spec} we compare the \aco\ line to the higher-$J$ CO lines detected in 2 and 3\,mm with NOEMA \citep{Neri2020}. Specifically we compare the FWHM from the Gaussian fit to the \aco, with the average of the FWHM from the respective Gaussian fits to the higher-$J$ transitions \citep[expressed as $\Delta$V in][]{Neri2020}. To quantify the difference, we use the relative difference of the two, expressed as $\rm |FWHM-\Delta V|/\Delta V$.  We find that the line profiles are similar for most sources in our sample, with 12/14 sources having a \aco\ FWHM consistent within 50\% of the $\Delta$V from \cite{Neri2020} for the higher $J$ transitions. The two sources with a significant difference in line-widths are HerBS-34 with a ratio of FWHM/$\Delta$V$=1.8\pm0.3$, and HerBS-58 with a ratio of FWHM/$\Delta$V$=0.4\pm0.1$. The large ratio of HerBS-34 is most likely due to the low signal to noise ratio of the \aco\ observations. The case of HerBS-58 is more complex. When comparing to the previously observed \cco\ and $\rm [CI](^3P_1-^3P_0)$ in Fig.~\ref{fig:herbs58} it becomes apparent that we are only detecting part of the line emission in \aco. Specifically, we are only detecting the red component of the line. Consequently, as we are only fitting the red component, this also leads to an underestimation of the error on the FWHM/$\Delta$V. When comparing the \aco\ moment-0 map to the resolved velocity map of $\rm [CI](^3P_1-^3P_0)$ in \citet{Neri2020}, we find that the \aco\ is located in the red-shifted western region, with no emission covering the blue-shifted eastern part of the [CI] emission, consistent with the lack of emission seen for the expected blue component of the line. A more detailed analysis of the CO and [CI] emission in HerBS-58, will be provided in Ismail et al. (in prep.) based on new high-angular resolution NOEMA data. 

The resolution of $\sim 1\arcsec$ achieved with the VLA has allowed for all sources to be at least partially resolved (see Table~\ref{tab:sources}, and Fig.~\ref{fig:spec}). Half of the sample shows compact to somewhat extended emission over 9--18\,kpc scales. For the other 7/14 sources extended and complex structure is revealed, over scales of 18--25\,kpc, with HerBS-54, 58, 79, 89a, 113, and 154 showing multiple peaks in their emission, while HerBS-95W has a single peak and extends over 20\,kpc. The complex, multi-peak structure observed for these sources is a possible indicator for lensing, as shown in \cite{Berta2021} for HerBS-89a. However, we are not able to perform a detailed lensing analysis in order to confirm this with the current data. A brief discussion on this point is provided in Appendix~\ref{app:lensing}.

\subsection{Radio continuum}\label{sec:continuum_results} 

As reported in Sect.~\ref{sec:obs}, due to the limited sensitivity of the VLA data, the majority of the sample is not detected in continuum, with the exception of four sources. The typical rms levels are between $\sim 4-45$\,$\mu$Jy\,beam$^{-1}$ in the observed frequency range of 20 to 38\,GHz. In Table~\ref{tab:cont} we give the continuum properties of the full sample. The four sources detected in the high-frequency radio continuum are: HerBS-43a, HerBS-43b, HerBS-54, and HerBS-70E; the continuum emission is displayed as white contours in the respective panels in Fig.~\ref{fig:spec}. 

Sources HerBS-43a and HerBS-43b are similarly compact in the 22.2\,GHz (rest-frame 93.5\,GHz and 112.2\,GHz, respectively) continuum than in the \aco\ emission, but we observe a small offset of $\sim 0\farcs5$ between the continuum and \aco\ emission for both sources. Similarly, the continuum of HerBS-70E, detected at 27.1 and 34.8\,GHz (rest-frame 89.6\,GHz and 115.1\,GHz) is also somewhat offset with respect to the \aco\ emission, and displays a more compact morphology. Interestingly, HerBS-54, detected at 33.0\,GHz (rest-frame 113.6\,GHz) has continuum emission that is aligned with only the south peak of the \aco\ emission, although there is some extension towards the north. This could be a result of the north component being fainter in the continuum and therefore not detected in these observations. 

For the sources detected in the radio continuum we define simple SED models, in which we use the modified black body (MBB) fits done to the NOEMA continuum data in \citet{Neri2020}, with correction for the cosmic microwave background (CMB) following \cite{daCunha2013}, and simply add a power-law spectrum of $S_\nu \propto \nu^{-0.8}$ representing the synchrotron emission due to star formation (SF synchrotron), normalized on the basis of the radio/far-infrared correlation taking into account the redshift evolution described in \citet{Delhaize2017} and the infrared emission derived using the DL07 model \citep{DraineLi2007} as reported in \citep{Neri2020} (see Table~\ref{tab:Sample}).
We show these SEDs in Fig.~\ref{fig:radio-continuum}. 

In the case of HerBS-43a, the continuum at 22.2\,GHz lies on the expected SF synchrotron power-law, while the upper limits at 27.2 and 37.4~GHz are also consistent. Although they lie above the expected emission based on the \citet{Delhaize2017} relation, HerBS-43b and HerBS-54 have flux densities that are compatible with the synchrotron emission due to star formation as they are in agreement with expectations based on the \citet{Magnelli2015} relation. 
 However, we note that measurements at lower frequencies (e.g., 4-8~GHz) would be useful to further assess the slope and derive the properties of the radio emission in these SMGs and, in particular, to estimate the contribution of the free-free (Bremsstrahlung) emission with a power-law radio spectrum of $S_\nu \propto \nu^{-0.1}$ that contributes significantly at higher frequencies \cite[e.g.,][]{Condon1992,Thomson2014}. All the sources with only upper limits in the radio continuum are compatible with the synchrotron emission levels derived from the radio/far-infrared relationship.
    
In the case of HerBS-70E, the flux densities are well above the expected level of the synchrotron emission due to star formation. This clearly indicates the presence of a radio luminous AGN in this source (see Fig.~\ref{fig:radio-continuum}). HerBS-70E was also detected in the FIRST survey \citep{Becker1995} at 1.4~GHz and with LOFAR at $\rm 126-173 \, MHz$ \citep{Hardcastle2016}, tracing the increase of the radio flux density at these lower frequencies (corresponding to 4.63~GHz and 416-572\,MHz in the rest-frame), including the expected turn-over of the radio AGN emission. Based on the 1.4\,GHz data we estimate a radio luminosity of $L_{\rm 1.4GHz} = 2.4\times10^{26}$\,W\,Hz$^{-1}$, placing it among typical radio luminous AGN at these redshifts. A more detailed analysis of this source will be presented in a separate paper.

\section{Discussion}\label{sec:discussion}

In this discussion, we use the observed \aco\ line properties to determine the molecular gas properties of our sample of luminous DSFGs, namely: the CO luminosities and masses (Sect.~\ref{sec:mgas}), the gas properties, including the gas depletion time, that are compared to other samples previously observed in \aco\ (Sect.~\ref{sec:tdep}), and the gas to dust mass ratios (Sect.~\ref{sec:gdr}). 
In Sect.~\ref{sec:coratios}, we estimate the CO line ratios by combining the \aco\ measurements with the previously observed higher-$J$ CO transitions from \citet{Neri2020}, and, in  Sect.~\ref{sec:sleds}, we analyse the spectral line energy distributions (SLEDs) of our sample with radiative transfer modeling. An important caveat that must be noted, and could affect most if not all of the points in this discussion, is the likelihood of our observations missing extended \aco\ emission. This would increase the line fluxes and luminosities, but it is not clear by how much, and how significant that difference could be.

\subsection{CO luminosities, molecular gas masses, and the choice of $\alpha_{CO}$}
\label{sec:mgas}

The direct measure of the \aco\ emission allows for the estimation of the line luminosity, $L^\prime_{\rm CO(1-0)}$, and molecular gas masses, $\rm M_{\rm H_2}$, without the uncertainty of conversion from higher-$J$ CO transitions. We estimate these values following the standard equations given below \cite[e.g.,][]{Carilli-Walter2013}: 
\begin{equation*}\label{eq:Lprimeco}
	L^\prime_{\rm CO} = 3.25 \times 10^7 \times \frac{S_{\rm CO}\,\Delta \upsilon_{CO}\,D_{\rm L}^2}{\nu_{CO,rest}^{2}\times(1+z)}\,\rm [K\,km\,s^{-1}\,pc^{-2}]
\end{equation*}

\begin{equation*}\label{eq:gas-mass}
	M_{\rm H_2} = \alpha_{CO} \times L^\prime_{\rm CO(1-0)} \,[M_\sun]
\end{equation*}

\noindent where $S_{\rm CO}\,\Delta v_{CO}$ is the measured flux of the line in Jy\,km\,s$^{-1}$, $D_{\rm L}$ is the luminosity distance in Mpc, $\nu_{CO,rest}$ is the rest frequency of the line in GHz, and $\rm \alpha_{CO}$ is the $\rm CO(1-0)-H_2$ conversion factor. 

The estimation of $M_{\rm H_2}$ remains uncertain even with the direct measure of the \aco\ emission, due to the dependence on $\rm \alpha_{CO}$, which itself is dependent on other properties of the galaxies, such as metallicity, that are difficult to determine over a large sample of galaxies, especially at high redshifts \citep[see e.g.][]{Dunne2021,Bolatto2013}. Historically, studies of classical sub-mm galaxies (SMGs) have assumed an $\rm \alpha_{CO} = 0.8\,\rm M_\sun \, {\rm (K \, km \, s^{-1} \, pc^2)^{-1}}$ \citep[see review by][]{Carilli-Walter2013}, in agreement with the dynamical constraints on the  $\rm \alpha_{CO}$ in local starburst galaxies placing it within the range of $0.8 - 1.5\,\rm M_\sun \, {\rm (K \, km \, s^{-1} \, pc^2)^{-1}}$ \cite[e.g.,][]{DownesSolomon1998,Genzel2010}. 
Normal star-forming galaxies seem to have an  $\rm \alpha_{CO} \sim 4\,\rm M_\sun \, {\rm (K \, km \, s^{-1} \, pc^2)^{-1}}$ \citep[e.g.][]{Tacconi2013, Tacconi2018, Genzel2015, Carilli-Walter2013} consistent with Galactic studies \citep[e.g.][]{Bolatto2013}. More recently a study on {\it Herschel} selected galaxies using multiple gas tracers to calibrate the gas mass, found an average $\rm \alpha_{CO} = 3 \,\rm M_\sun \, {\rm (K \, km \, s^{-1} \, pc^2)^{-1}}$ \citep{Dunne2021}. For the purposes of this discussion, we chose to use the results of \cite{Dunne2021}, and assume an $\rm \alpha_{CO} = 3 \,\rm M_\sun \, {\rm (K \, km \, s^{-1} \, pc^2)^{-1}}$, but in addition we examine the range of results possible when assuming the two extremes of $\rm \alpha_{CO} = 0.8 - 4.3 \,\rm M_\sun \, {\rm (K \, km \, s^{-1} \, pc^2)^{-1}}$. 
We note that during the final stages of this study a more updated analysis on the $\rm \alpha_{CO}$ was presented in \cite{Dunne2022}, using a much larger sample of 407 galaxies spanning up to $z\sim6$, and found $\rm \alpha_{CO} $ values consistent with those of \cite{Dunne2021}, assumed in our analysis. We also note that in a detailed high-resolution study of a high-$z$ luminous {\it Herschel} selected galaxy, \cite{Dye2022} estimated the gas mass of the galaxy following different tracers and methods - including from \aco\ assuming $\rm \alpha_{CO} = 3 \,\rm M_\sun \, {\rm (K \, km \, s^{-1} \, pc^2)^{-1}}$ - and found a surprising agreement between the resulting gas masses.

The estimated \aco\ luminosities and gas masses for our sample are given in Table~\ref{tab:COlum}. We find molecular gas masses of $1.2-13.2\times10^{11}$M$_\odot$, with no corrections made for the possible magnification due to gravitational lensing. The sources with the largest values ($\geq5.4\times10^{11}$) are all identified to have extended and complex structure and are likely magnified. For HerBS-89a, 
correction for gravitational lensing ($\mu = 5$) reduces the molecular gas mass from $(13.2\pm2.7)\times10^{11}$ to $(2.6\pm0.5)\times10^{11}$\,M$_\odot$ \citep[see also][]{Berta2021}. The large molecular gas masses measured are consistent with the necessary conditions to support the SFRs of such galaxies, and are in agreement with literature results for luminous DSFGs \citep[e.g.,][]{Tacconi2008,Ivison2011,Bothwell2013,Aravena2016,Harrington2021}.

\begin{table} % [th]
\begin{center}
\caption{Derived properties of the molecular gas}\label{tab:COlum}
\begin{tabular}{lcccc}
\hline 
\hline 
Source  &  ($\mu$)$L^\prime_{\rm CO}$ &  ($\mu$)M$_{\rm H_2}$ &  t$_{\rm dep}$ \\ 
&   [10$^{11}$ K\,km\,s$^{-1}$\,pc$^2$] & [10$^{11}$\,M$_\odot$] & [10$^{2}$\,Myr] \\ 
\hline 
HerBS-34  & $1.5\pm0.4$ & $4.5\pm1.2$ & $1.1\pm0.3$  \\ 
HerBS-43a & $1.7\pm0.5$ & $5.1\pm1.5$ & $1.5\pm0.5$  \\ 
HerBS-43b  & $0.4\pm0.2$ & $1.2\pm0.6$ & $0.8\pm0.4$  \\ 
HerBS-44  & $1.5\pm0.3$ & $4.5\pm0.9$ & $0.4\pm0.1$  \\ 
HerBS-54  & $2.7\pm0.6$ & $8.1\pm1.8$ & $3.4\pm0.8$  \\ 
HerBS-58  & $1.8\pm0.4$ & $5.4\pm1.2$ & $3.1\pm0.7$  \\ 
HerBS-70E  & $0.9\pm0.2$ & $2.7\pm0.6$ & $0.7\pm0.2$  \\ 
HerBS-70W  & $0.5\pm0.1$ & $1.5\pm0.3$ & $1.7\pm0.9$  \\ 
HerBS-79  & $2.3\pm0.5$ & $6.9\pm1.5$ & $3.8\pm0.9$  \\ 
HerBS-89a  & $4.4\pm0.9$ & $13.2\pm2.7$ & $4.6\pm1.0$  \\ 
HerBS-95E  & $0.6\pm0.2$ & $1.8\pm0.6$ & $1.6\pm0.6$  \\ 
HerBS-95W  & $2.2\pm0.4$ & $6.6\pm1.2$ & $4.1\pm0.9$  \\ 
HerBS-113 & $2.9\pm0.6$ & $8.7\pm1.8$ & $2.9\pm0.7$  \\ 
HerBS-154  & $2.3\pm0.4$ & $6.9\pm1.2$ & $0.9\pm0.2$  \\ 
\hline
%mean & 7.58 & 1.58 & 4.75 & 2.17 \\
%median & 7.5 & 1.6 & 4.9 & 1.58 \\
%\hline
\end{tabular}
\tablenotetext{}{{\bf Note.} Where $\mu$ is the magnification factor due to gravitational lensing, $L^\prime_{\rm CO}$ is the \aco\ luminosity, M$_{\rm H_2}$ is the molecular gas mass, and t$_{\rm dep}$ is the depletion time.}	
\end{center}
\end{table}

\begin{figure*}
\begin{center}
\includegraphics[width=0.49\textwidth]{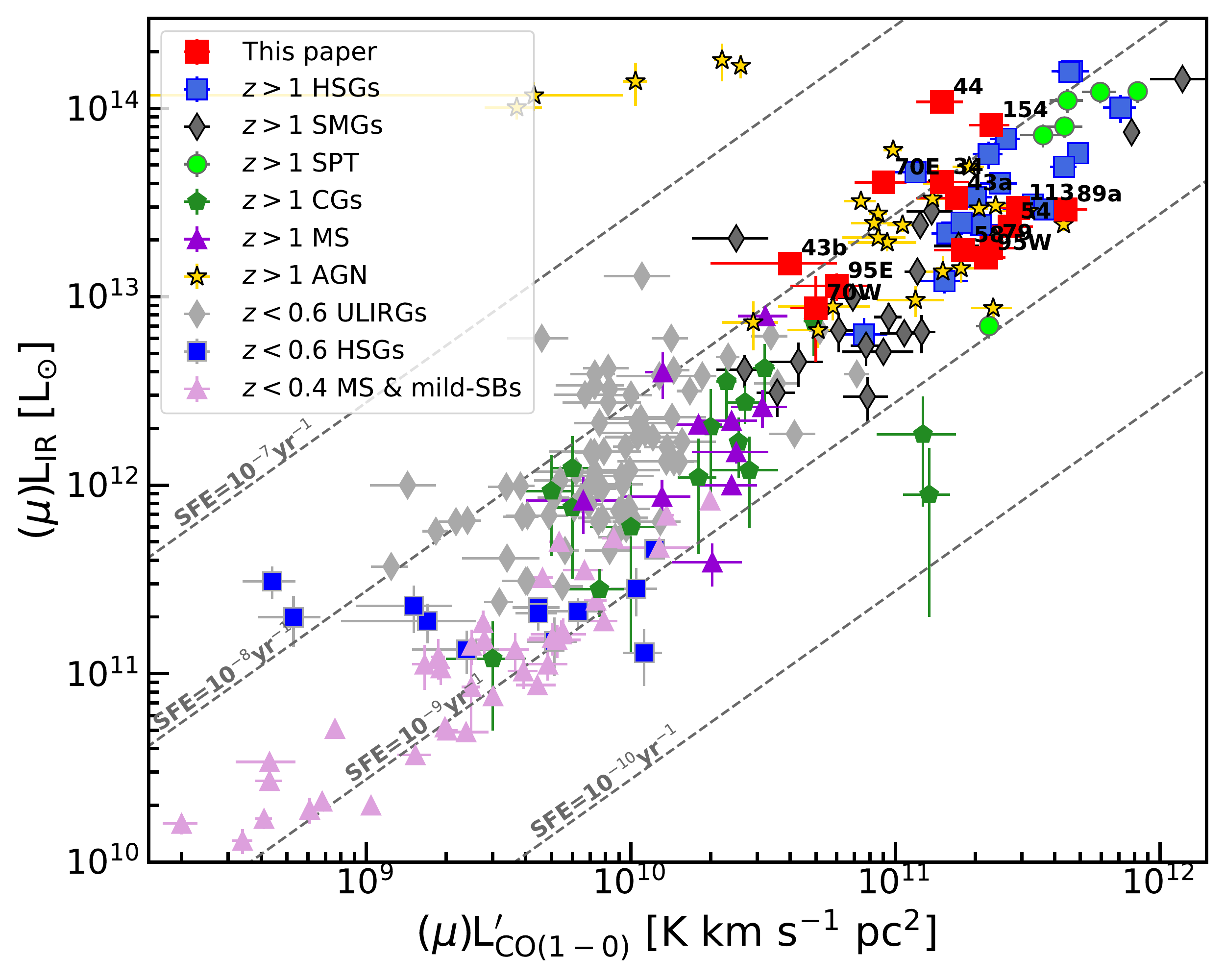}
\includegraphics[width=0.49\textwidth]{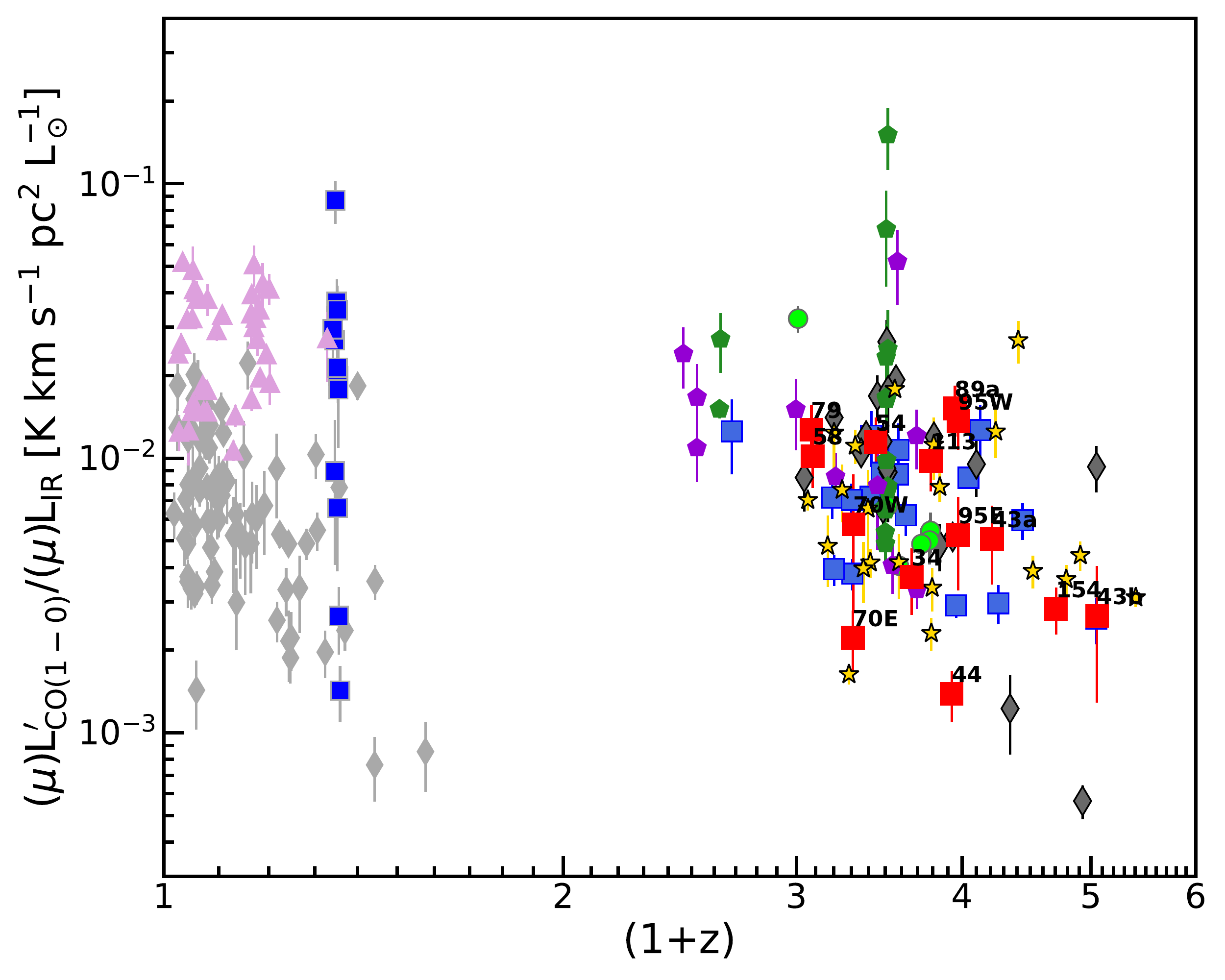}
\caption{$\it (left)$ Infrared luminosity (L$_\mathrm{IR}$; 8--1000$\mum$) as a function of L$^{\prime}_\mathrm{CO(1-0)}$ for the  DSFGs reported in this paper (shown as red squares and identified by their names).  For comparison we also plot low and high redshift DSFGs from the literature, categorized as sub-mm galaxies (SMGs), {\it Herschel} selected galaxies (HSGs), galaxies from the SPT survey, cluster galaxies (CGs), galaxies on the main sequence (MS), DSFGs hosting AGN (AGN) and ULIRGs (see section \ref{sec:tdep} for a detailed list of references).  Only sources with reported \aco\ emission lines have been included to avoid uncertainties in estimating the \aco\ luminosities from higher-$J$ transitions. Trends of constant SFE values are plotted with grey dashed lines. $\it (right)$  The L$^{\prime}_\mathrm{CO(1-0)}$/L$_{\rm IR}$ ratio as a function of ($1+z$). In both plots, corrections for amplification were not applied to the infrared and CO luminosities for the gravitationally amplified galaxies. 
} 
\label{fig:lirlco}
\end{center}
\end{figure*}

\subsection{Molecular Gas Depletion Times \& Star Formation Efficiencies}\label{sec:tdep}
Two parameters of interest when investigating the molecular gas properties of galaxies, are the star formation efficiency (SFE) and its inverse, the gas depletion time (t$_{\rm dep}$). The SFE  is a measure of the efficiency of star formation given the molecular gas reservoir available, and can be expressed as $\rm SFE = SFR/{\it M_{\rm gas}}$, where SFR[\,M$_{\odot}$\,yr$^{-1}$]$ = 1.09 \times 10^{-10} \, L_{\rm IR}$[L$_{\odot}$], and $L_{\rm IR}$ is the infrared luminosity at rest-frame 8--1000$\mu m$  \citep[][corrected for a \citealt{Chabrier2003} initial mass function]{Kennicutt1989}. The t$_{\rm dep}$ is a measure of the time it would take for the molecular gas to be depleted by the current SFR, assuming that the SFR will remain constant throughout this period and that no other processes (such as gas accretion or outflows) affect the gas reservoir. 

 To compare the line emission properties of our sample to previously observed DSFGs at low and high redshifts, we plot $L_{\rm IR}$ as a function of $L^\prime_{\rm CO(1-0)}$ and the ratio of  $L^\prime_{\rm CO(1-0)}/L_{\rm IR}$ in Figure~\ref{fig:lirlco}. For this comparison we only include DSFGs with measured \aco\, and we split the sources based on redshift and in the following categories: classical SMGs selected at 850$\mum$ and/or 1200$\mum$ \citep[see compilation in][]{Carilli-Walter2013}, with \aco\ measurements from \cite{Riechers2011c,Riechers2011e,Swinbank2011,Dannerbauer2019,Lestrade2011,Harris2010,Carilli2010,Ivison2011,Thomson2012,Sharon2016}; galaxies from the south pole telescope (SPT) survey \citep[e.g.][]{Weiss2013,Spilker2014}, and with \aco\ measurements presented in \cite{Aravena2013,Aravena2016}; {\it Herschel} selected galaxies (HSGs) from surveys such as Herschel-ATLAS \citep[e.g.][]{Maddox2018,Valiante2016} and the HerBS catalogue \citep{Bakx2018}, with \aco\ measurements from \cite{Harris2012,George2013,Bakx2020b}; galaxies from cluster studies \citep{Rudnick2017, Wang2018}; galaxies on the main sequence (MS) \citep{Aravena2014, Villanueva2017}; high-$z$ DSFGs hosting AGN \citep[from the compilations of][]{Harris2012, Carilli-Walter2013, Sharon2016, Penney2020}; and ULIRGs \citep{Solomon1997, Chung2009,Combes2011}. 

In Figure~\ref{fig:lirlco} we can see that our sample falls within the scatter of high redshift DSFGs, without much distinction between the different selection methods. Furthermore, although there is an apparent $L_{\rm IR}$-- $L^\prime_{\rm CO(1-0)}$ correlation for most sources, it is worth noting that the scatter corresponds to a factor of $\sim$10 in SFEs. When taking the ratio of $L^\prime_{\rm CO(1-0)}/L_{\rm IR}$ with redshift the comparison amongst samples becomes somewhat clearer as the effects of magnification are removed. The luminosity ratios of high redshift SMGs, HSGs, AGN, and SPT galaxies, cover the same range as lower redshift ULIRGs, while there is quite some overlap between these high redshift luminous DSFGs and MS galaxies at those redshifts.

Although, the ratios of $L^\prime_{\rm CO(1-0)}/L_{\rm IR}$ (or reverse) can be interpreted as a proxy to $t_{\rm dep}$ (or SFE) allowing for comparisons without the interference of $\rm \alpha_{CO}$ assumptions, such comparisons, especially between different populations, remain limited precisely due to the fact that the $\rm \alpha_{CO}$ could vary significantly among the range of the sources examined.

\begin{figure}
\centering
\includegraphics[width=0.48\textwidth]{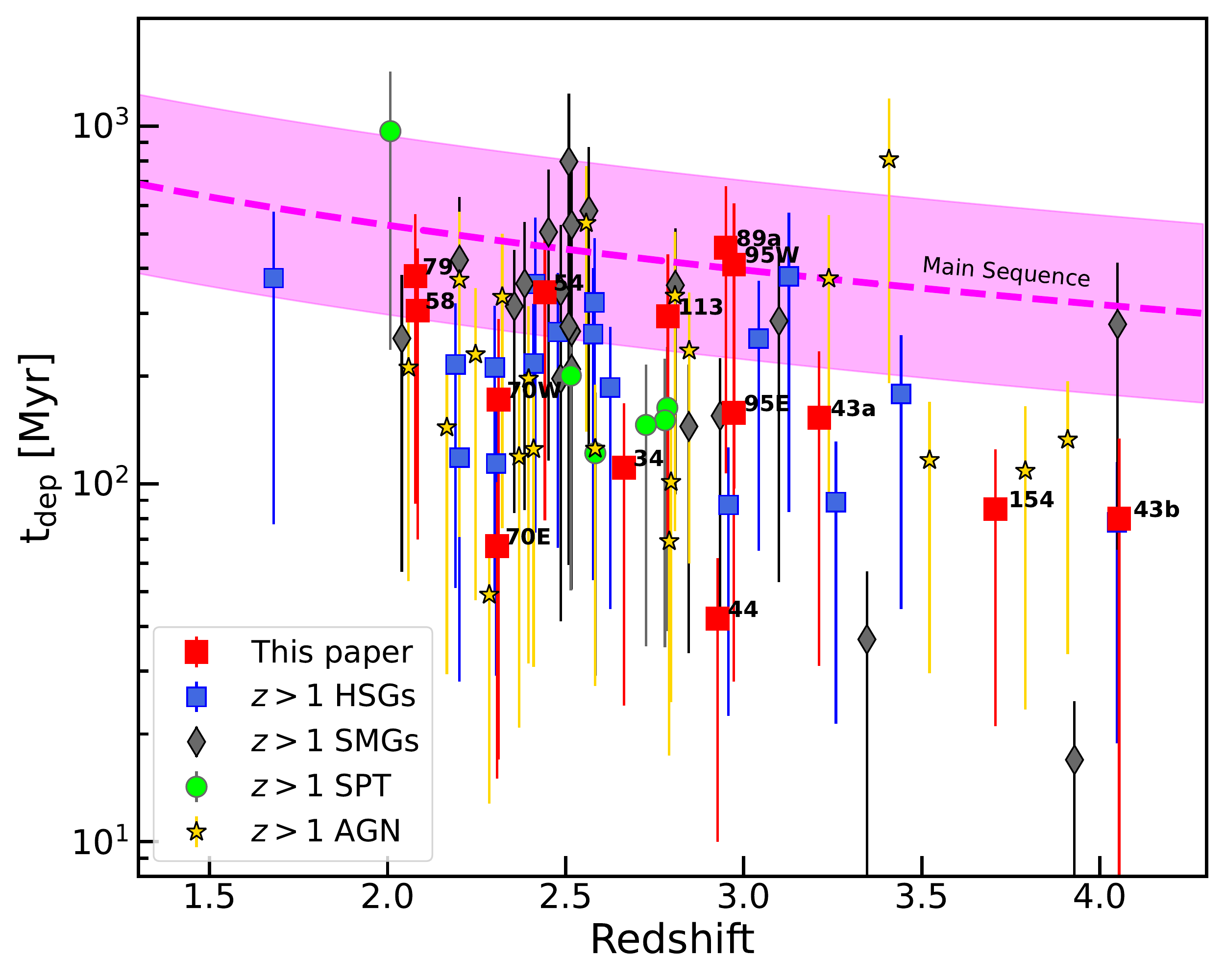}\\
\caption{Depletion time (t$_{\rm dep}$) as a function of redshift. Our sample is plotted with red squares and the individual sources are labelled. We compare to sources from the literature, specifically the high redshift DSFGs detected in \aco\ (see relevant references in section~\ref{sec:tdep}). We estimate the t$_{\rm dep}$ of the literature sources making the same assumptions as for our sample, with an $\alpha_{\rm CO} = 3 \,\rm M_\sun \, {\rm (K \, km \, s^{-1} \, pc^2)^{-1}}$. However, the error bars plotted, include the range of t$_{\rm dep}$ that would be estimated using $\alpha_{\rm CO}$ from 0.8 to 4.3$\,\rm M_\sun \, {\rm (K \, km \, s^{-1} \, pc^2)^{-1}}$}\label{fig:tdep}
\end{figure}

 To further examine the t$_{\rm dep}$ of our sample we plot in Figure~\ref{fig:tdep} t$_{\rm dep}$ as a function of redshift, comparing to the high redshift luminous DSFGs for which it is more likely that a similar value of $\alpha_{\rm CO}$ could apply. We also plot the range of t$_{\rm dep}$ covered by MS galaxies from the relation presented in the \cite{Tacconi20} review. To examine the effect that the assumed $\alpha_{\rm CO}$ can have on the t$_{\rm dep}$  values, we include in the plotted errorbars the range of t$_{\rm dep}$ that results from assuming $\alpha_{\rm CO}=0.8\,\rm and\, 4.3$.

The range in 
t$_{\rm dep}$ that we find for our sample is consistent with what is seen for the luminous DSFGs 
at the same redshifts. Furthermore, we observed the decreasing trend of t$_{\rm dep}$ with redshift that has been previously observed for the general population of DSFGs \citep[see review by][]{Tacconi20}.  

 Overall, we find a t$_{\rm dep}$ of 40--460\,Myr for our sources (see Table~\ref{tab:COlum}). These values place our sample both on and below the MS values for the redshift range covered, demonstrating that luminous DSFGs from {\it Herchel} selected samples, include both normal and star-bursting galaxies \citep[see also][]{Berta2021}.

\subsection{Gas to Dust ratios}\label{sec:gdr}

\begin{figure}
\centering
\includegraphics[width=0.48\textwidth]{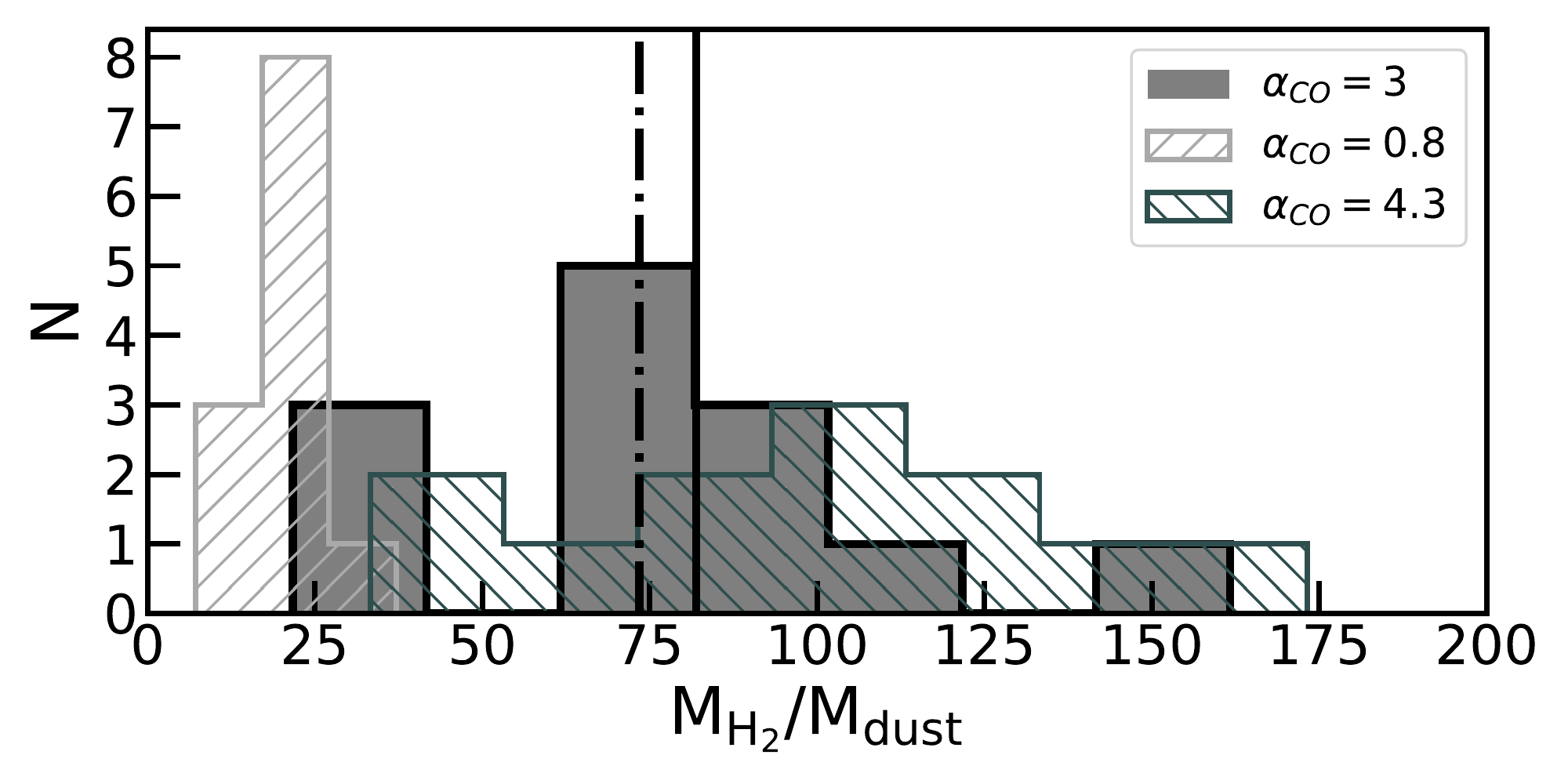}
\caption{Histogram of the gas to dust ratio (M$_{\rm gas}$/M$_{\rm dust}$) of our sample, for an  $\alpha_{\rm CO} = 3 \,\rm M_\sun \, {\rm (K \, km \, s^{-1} \, pc^2)^{-1}}$, is plotted in solid. We also plot the histograms that result from assumptions of  $\alpha_{\rm CO} = 0.8$ and 4.3$\,\rm M_\sun \, {\rm (K \, km \, s^{-1} \, pc^2)^{-1}}$ for comparison.} \label{fig:gasdust}
\end{figure}

We use the dust masses (M$_{\rm dust}$) derived for our sample by \citet{Neri2020} using 
\citet{DraineLi2007} templates (see Table~\ref{tab:Sample}), to estimate the gas to dust ratio ($\rm M_{\rm H_2}$/M$_{\rm dust}$). 
We find a wide range of values, $\rm M_{\rm H_2}/M_{\rm dust} = 27 - 167$, with a mean of $\sim82$  (see solid histogram in Figure~\ref{fig:gasdust}). For comparison, $z\sim0.3$ {\it Herschel} selected galaxies have ratios of 64--261, with a mean of $128^{+54}_{-35}$ \citep{Dunne2021}, these values remain consistent with the larger sample of DSFGs of \cite{Dunne2022} at $z<6$. For nearby star-forming galaxies the ratio is $\sim70$ \citep[][]{Sandstrom2013}. Although, the majority of our sample is in agreement with previous measurements of the gas to dust ratio, HerBS-43b, HerBS-95E, and HerBS-34 have surprisingly low ratios of 27, 36, and 39, respectively. These low ratios could be a result of variation of $\alpha_{CO}$ amongst the sample, or due to the presence of an AGN contaminating the SED and artificially elevating the consequent M$_{\rm dust}$ estimates. Indeed in the study of \cite{Sharon2016} where the properties and line ratios of AGN and SMGs were compared, they found that although there is significant overlap amongst the two samples the lowest values were those of AGN.

In Figure~\ref{fig:gasdust} we also plot the distribution of the ratio when assuming $\alpha_{\rm CO}=0.8\,\rm and\, 4.3$ with hollow histograms. An assumption of $\alpha_{\rm CO}=0.8 $ results in the whole sample having ratios below 40, while an  $\alpha_{\rm CO}=4.3$ does not change the distribution significantly.

\subsection{CO line ratios \& SLEDs}\label{sec:coratios}

\begin{figure}
\centering
\includegraphics[width=0.48\textwidth]{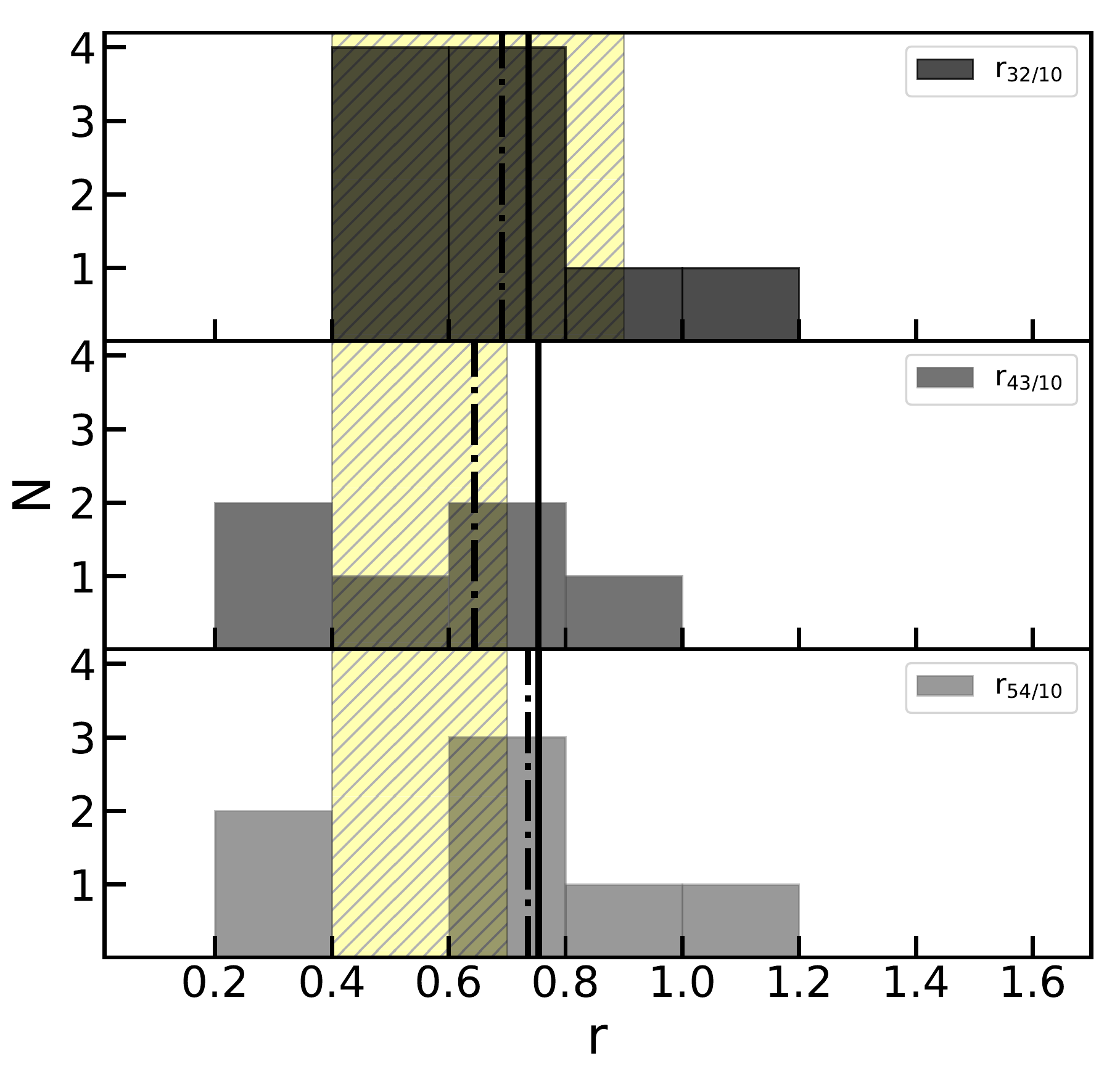}
\caption{Histograms of the CO brightness temperature ratios, for different transitions: (top) r$_{32/10} = \rm L^\prime_{CO(3-2)}/L^\prime_{CO(1-0)}$, (middle) r$_{43/10} = \rm L^\prime_{CO(4-3)}/L^\prime_{CO(1-0)}$, (bottom) r$_{54/10} = \rm L^\prime_{CO(5-4)}/L^\prime_{CO(1-0)}$. The solid line corresponds to the mean of the distribution, while the dot-dashed corresponds to the median. The highlighted area corresponds to the range of average ratios found in the literature for the corresponding transitions. } \label{fig:rcohist}
\end{figure}

\begin{figure}
\centering
\includegraphics[width=0.5\textwidth]{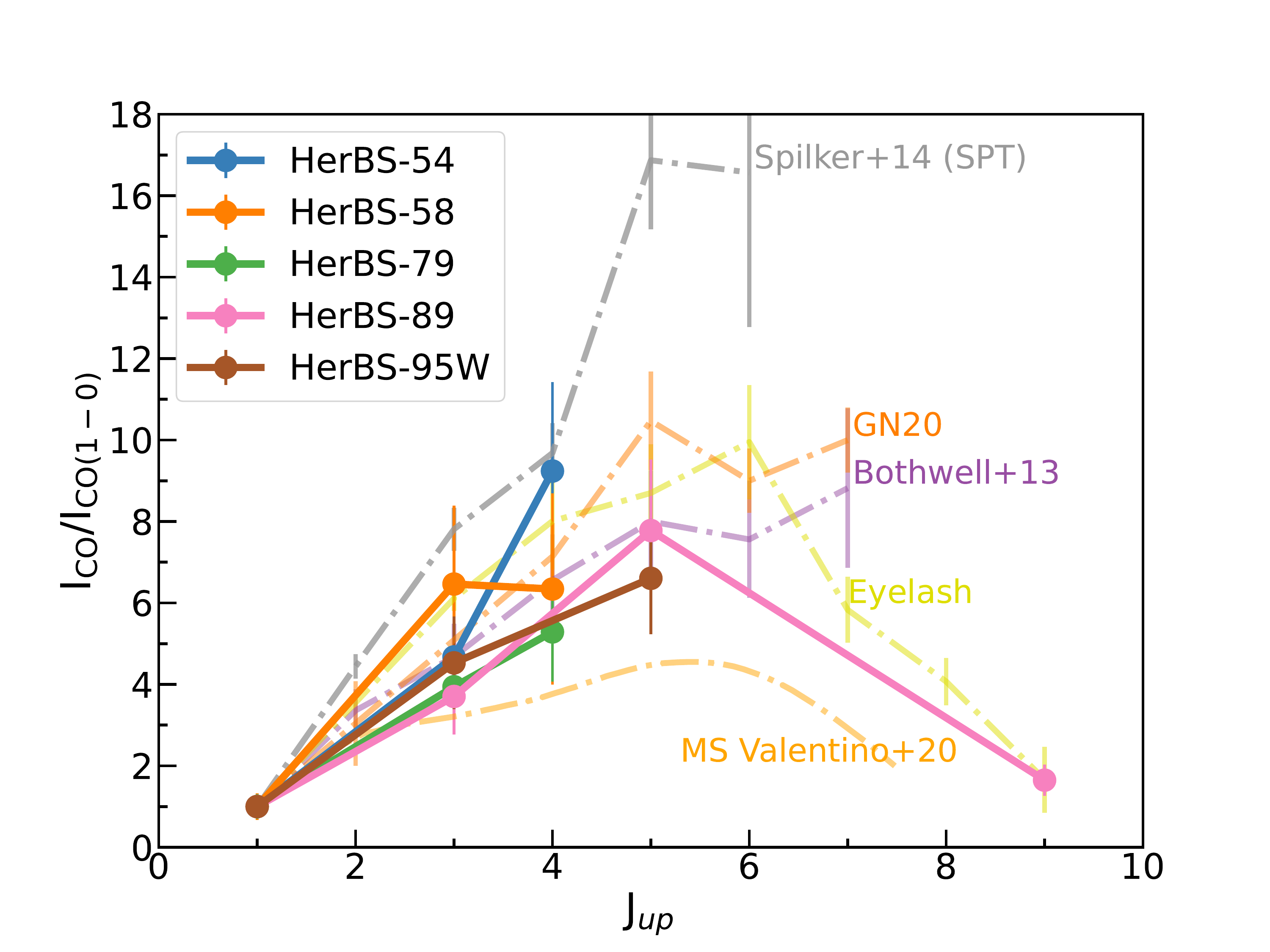}\\
\includegraphics[width=0.5\textwidth]{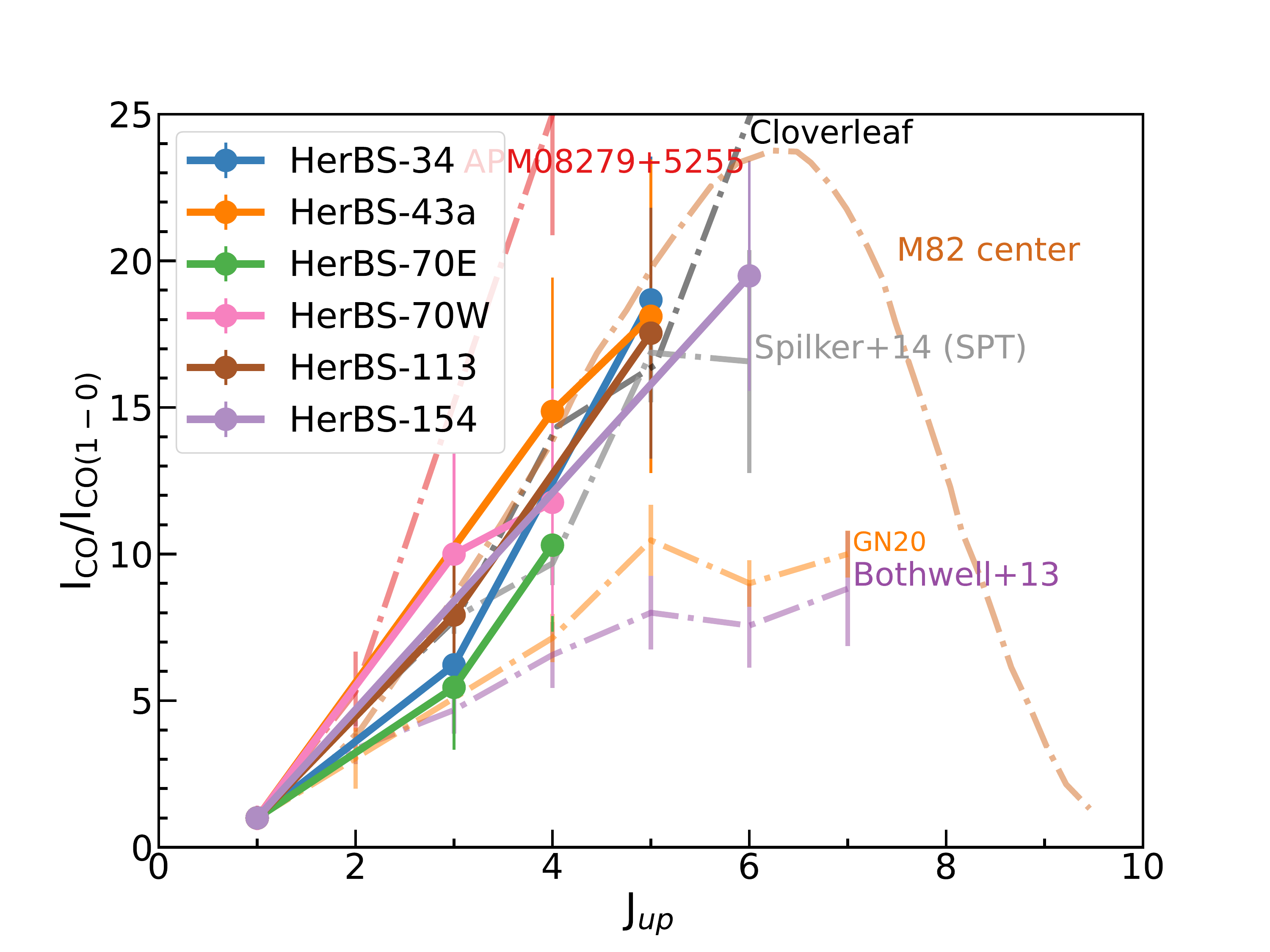}\\
\includegraphics[width=0.5\textwidth]{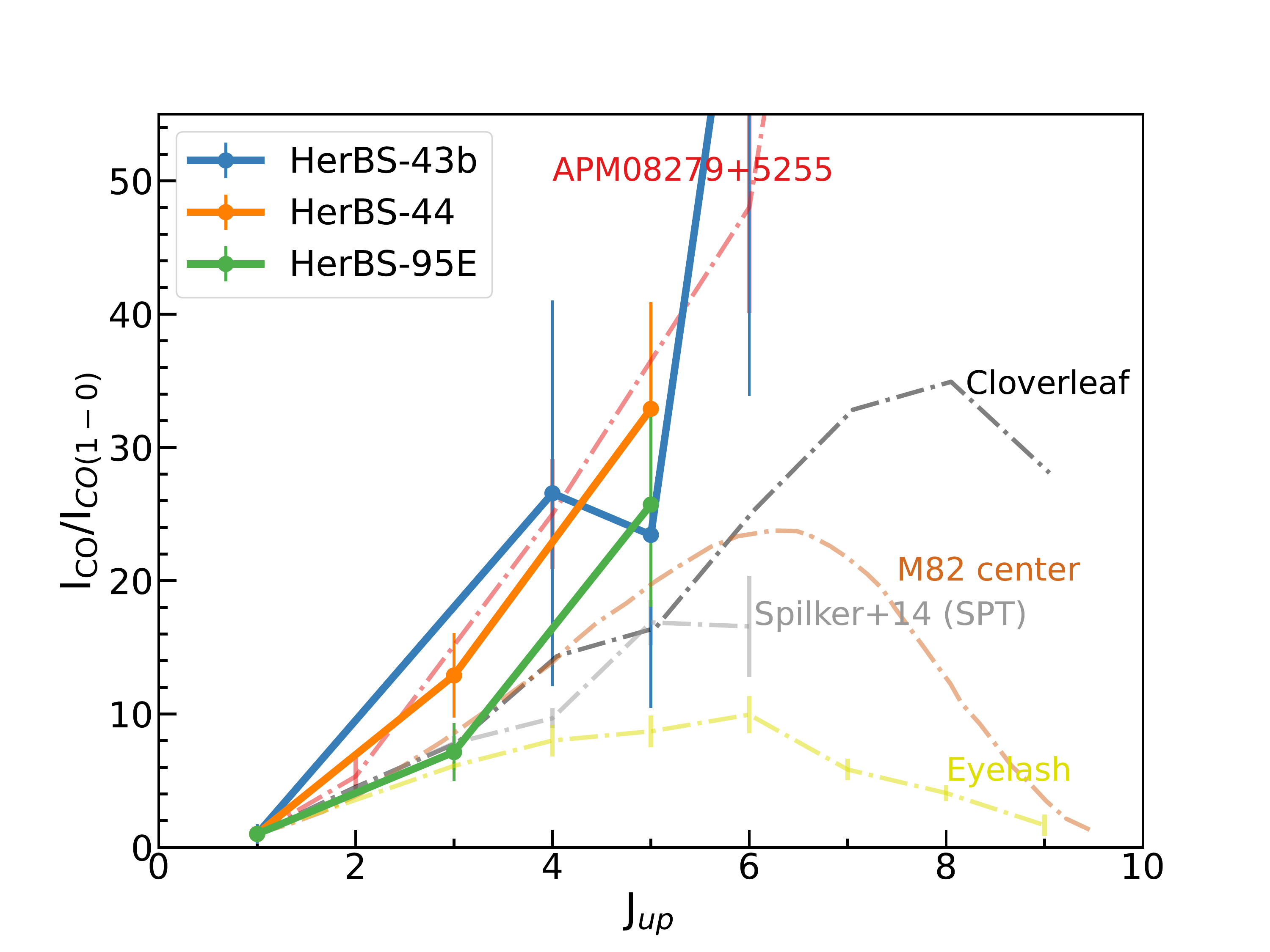}\\
\caption{CO SLEDs, normalized to \aco\ for our sample. The sample has been split in three sub-samples for easier distinction of the sources in the figures. The normalized integrated CO fluxes measured for each source are plotted as circles. For comparison we also plot SLEDs of other sources from the literature in dot-dashed curves (see section~\ref{sec:sleds} for references). Note that for HerBS-89a we include the CO(9$-$8) transition and use the best fit SLED model originally presented in \citet{Berta2021}. }\label{fig:cosled}
\end{figure}

As our sample has been observed in multiple CO transitions in addition to \aco, 
we can directly measure the CO brightness temperature ratios, $r$, by taking the ratio of the L$^\prime_{CO}$ of the different transitions over \aco\ (e.g., $r_{32/10} = \rm L^\prime_{CO(3-2)}/L^\prime_{CO(1-0)}$). For HerBS-58 that only has the red component of the higher-$J$ transitions detected in \aco\, we calculate an upper limit on the total flux and luminosity of the \aco\ line over the full linewidth observed for the higher-$J$ lines and report the corresponding lower limit on the $r$ values of this source. We find that the line ratios have a similar range of values across the different CO transitions, between 0.3--1.7 (see Table~\ref{tab:ratios}). In Figure \ref{fig:rcohist} we plot the observed distributions for $r_{32/10} = \rm L^\prime_{CO(3-2)}/L^\prime_{CO(1-0)}$, r$_{43/10} = \rm L^\prime_{CO(4-3)}/L^\prime_{CO(1-0)}$, r$_{54/10} = \rm L^\prime_{CO(5-4)}/L^\prime_{CO(1-0)}$, with the typical range of values found for SMGs shown with a yellow filled region. Typical values of $r$ for SMGs are $r_{32/10} =0.4-0.9$, $r_{43/10} =0.4-0.7$, $r_{54/10} =0.4-0.7$, and $r_{65/10} =0.2-0.5$ \citep[e.g.,][and references therein]{Ivison2011,Bothwell2013,Spilker2014,Sharon2016,Harrington2021,Carilli-Walter2013}. 
The majority of our sample have ratios consistent with the typical SMG values. However, it is interesting to note that the majority of the r$_{54/10}$ ratios are toward the higher end of the typical SMG values, although the small sample size limits the significance of this. HerBS-43b, and HerBS-44 have ratios more in line with starburst-quasar systems such as APM08279$+$5255 \citep[e.g.,][]{Riechers2009}, but are still within the range of possible values for SMGs \citep[see][]{Sharon2016}. 
We note that the $r_{65/10}$ of HerBS-43b is quite large, with a value that is nearly twice that of APM08279$+$5255; however, there is a large uncertainty on the CO(6$-$5) flux detected by \citet{Neri2020}, which is reflected in the large error on this ratio.

\begin{table}[th]
\begin{center}
\caption{CO brightness temperature ratios}\label{tab:ratios}
\begin{tabular}{lcccc}
\hline 
\hline 
Source  & r$_{32/10}$ & r$_{43/10}$ & r$_{54/10}$ & r$_{65/10}$\\ 
\hline 
HerBS-34 & $0.7\pm0.2$ & $-$ & $0.7\pm0.2$ & $-$  \\ 
HerBS-43a & $-$ & $0.9\pm0.3$ & $0.7\pm0.2$ & $-$  \\ 
HerBS-43b & $-$ & $1.7\pm0.9$ & $0.9\pm0.5$ & ($2.1\pm1.1$)  \\ 
HerBS-44 & $1.4\pm0.4$ & $-$ & $1.3\pm0.3$ & $-$  \\ 
HerBS-54 & $0.5\pm0.1$ & $0.6\pm0.1$ & $-$  & $-$  \\ 
HerBS-58 & $>0.4$ & $>0.2$ & $-$  & $-$  \\ 
HerBS-70E & $0.6\pm0.2$ & $0.6\pm0.2$ & $-$  & $-$  \\ 
HerBS-70W & $1.1\pm0.4$ & $0.7\pm0.2$ & $-$  & $-$  \\ 
HerBS-79 & $0.4\pm0.1$ & $0.3\pm0.1$ & $-$ & $-$  \\ 
HerBS-89a & $0.4\pm0.1$ & $-$ & $0.31\pm0.07$ & $-$  \\ 
HerBS-95E & $0.8\pm0.2$ & $-$ & $1.0\pm0.3$ & $-$  \\ 
HerBS-95W & $0.5\pm0.1$ & $-$ & $0.3\pm0.1$ & $-$   \\ 
HerBS-113 & $0.9\pm0.3$ & $-$ & $0.7\pm0.2$ & $-$   \\ 
HerBS-154 & $-$ & $-$ & $-$ & $0.5\pm0.1$  \\ 
\hline 
\hline
\end{tabular}
\label{table:CO-line-ratios}
\end{center}
\end{table}

In Figure~\ref{fig:cosled}, we plot the integrated CO fluxes of our sources, normalized to the \aco. We have  divided the sources into three different plots for easier comparison with the SLEDs of other DSFG samples
from \citet{Bothwell2013} (32 SMGs at $1.2 < z < 4.1$), and \citet{Spilker2014} (22 SPT-selected SMGs at $2.0 < z < 5.7$), and the mean SLED of ``main sequence'' star-forming galaxies from \citet{Valentino2020}. We also make a comparison with individual sources: Cosmic Eyelash \citep{Swinbank2011,Danielson2011}; GN20 \citep{Carilli2010,Cortzen2020}; the starburst-quasar galaxies Cloverleaf \citep{Barvainis1997,Weiss2003,Bradford2009, Riechers2011f}, and APM08279$+$5255 \citep{Papadopoulos2001,Riechers2006,Weiss2007,Riechers2009}; and M\,82 \citep{Weiss2005}. All SLEDs used for comparison are also normalised to the \aco. 

As can be seen in Figure~\ref{fig:cosled}, when comparing the normalised integrated flux values and respective curves, sources HerBS-54,-58,-79,-89a,-95W are most comparable to other high-$z$ SMGs, such as the Cosmic Eyelash, GN20, and the SMG sample of \citet{Bothwell2013}.
Sources HerBS-34,-43a,-70E,-70W,-113,-154 are most comparable to the SPT SMGs of \citet{Spilker2014}, the Cloverleaf galaxy, and the M82 center.
Finally, sources HerBS-43b,-44,-95E all lie between the two quasar-starburst systems Cloverleaf and APM08279$+$5255, which could indicate that these sources likely host an AGN. However, CO transitions at $J_{up}>6$ would be necessary to determine the presence of AGN excitation. Overall, the wide range of CO excitation we find for these sources demonstrates the importance of observing the \aco\ emission in galaxies when studying their molecular gas properties.

\begin{table*}[th]
\begin{center}
\caption{Properties derived from the radiative transfer analysis, where both the median (med) and the maximum likelihood (max) results for each property are given (with the exception of $P_{\rm thermal}$)}\label{tab:sledfit}
\begin{tabular}{lccccccccc}
\hline
\hline
Source & \multicolumn{2}{c}{log($n_{\rm H2}$)} & \multicolumn{2}{c}{log($T_{\rm K}$)} & \multicolumn{2}{c}{log($N_{\rm CO}$/dV)} & \multicolumn{2}{c}{log($P_{\rm thermal})$} \\
       & \multicolumn{2}{c}{[cm$^{-3}$]} & \multicolumn{2}{c}{[K]} & \multicolumn{2}{c}{[cm$^{-2}$\,km$^{-1}$\,s]} & \multicolumn{2}{c}{[K\,cm$^{-3}$]} \\
	& med & max & med & max & med & max & med   \\
\hline

HerBS-34 & 3.69$^{+0.95}_{-0.79}$ & 4.61 & 2.06$^{+0.42}_{-0.40}$ & 1.93 & 18.09$^{+0.28}_{-0.42}$ & 18.34 & 5.80$^{+0.71}_{-0.65}$ & \\
HerBS-43a & 2.97$^{+0.66}_{-0.57}$ & 2.66 & 2.22$^{+0.49}_{-0.49}$ & 2.02 & 17.58$^{+0.52}_{-0.65}$ & 17.87 & 5.29 $^{+0.45}_{-0.62}$ & \\
HerBS-43b & 3.92$^{+ 0.73}_{- 0.81}$ & 3.95 & 2.24$^{+ 0.39}_{- 0.45}$ & 2.36 & 17.82$^{+ 0.47}_{- 0.68}$ & 18.01 & 6.19$^{+ 0.54}_{- 0.72}$ & \\
HerBS-44 & 3.76$^{+0.73}_{-0.59}$ & 3.84 & 2.27$^{+0.34}_{-0.41}$ & 2.33 & 17.65$^{+0.57}_{-0.53}$ & 17.91 & 6.08$^{+0.47}_{-0.43}$ &  \\
HerBS-54 & 3.18$^{+1.00}_{-0.69}$ & 2.84 & 2.09$^{+0.44}_{-0.43}$ & 2.50 & 17.89$^{+0.41}_{-0.41}$ & 17.93 & 5.32$^{+0.75}_{-0.49}$ &  \\
HerBS-58 & 2.83$^{+0.72}_{-0.51}$ & 2.96 & 2.00$^{+0.61}_{-0.62}$ & 1.81 & 16.74$^{+0.68}_{-0.72}$ & 16.68 & 4.96$^{+0.41}_{-0.48}$ & \\
HerBS-70E & 2.86$^{+0.72}_{-0.52}$ & 3.38 & 2.25$^{+0.45}_{-0.44}$ & 2.39 & 17.57$^{+0.49}_{-0.60}$ & 17.37 & 5.19$^{+0.51}_{-0.52}$ & \\
HerBS-70W & 3.06$^{+0.74}_{-0.57}$ & 3.35 & 2.05$^{+0.56}_{-0.51}$ & 1.88 & 17.27$^{+0.67}_{-0.85}$ & 16.55 & 5.26$^{+0.48}_{-0.68}$ & \\
HerBS-79 & 2.49$^{+0.42}_{-0.33}$ & 2.18 & 2.29$^{+0.45}_{-0.48}$ & 2.40 & 16.89$^{+0.42}_{-0.51}$ & 17.01 & 4.82$^{+0.36}_{-0.37}$ & \\
HerBS-95E & 3.93$^{+0.66}_{-0.74}$ & 3.44 & 2.25$^{+0.33}_{-0.39}$ & 2.38 & 18.03$^{+0.33}_{-0.88}$ & 18.31 & 6.19$^{+0.56}_{-0.72}$ & \\
HerBS-95W & 2.55$^{+0.66}_{-0.74}$ & 2.21 & 2.31$^{+0.44}_{-0.50}$ & 2.65 & 17.21$^{+0.35}_{-0.37}$ & 17.23 & 4.93$^{+0.34}_{-0.38}$ & \\
HerBS-113 & 3.07$^{+0.83}_{-0.63}$ & 2.62 & 2.10$^{+0.51}_{-0.43}$ & 2.51 & 17.78$^{+0.44}_{-0.53}$ & 17.66 & 5.28$^{+0.59}_{-0.57}$ & \\ 
HerBS-154 & 3.09$^{+0.63}_{-0.64}$ & 3.16 & 2.24$^{+0.48}_{-0.47}$ & 1.87 & 17.51$^{+0.57}_{-0.92}$ & 17.78 & 5.44$^{+0.41}_{-0.64}$ & \\
\hline
\end{tabular}
\end{center}
\end{table*}

\subsection{Radiative Transfer Modelling}\label{sec:sleds}

Using the large velocity gradient (LVG) statistical equilibrium method \citep[e.g.,][]{Sobolev1960}, we modeled the molecular gas excitation conditions through the observed integrated fluxes of the multi-$J$ CO lines. We adopt the one-dimensional (1D) non-LTE radiative transfer code \texttt{RADEX} \citep{vanderTak2007}, with an escape probability of $\beta = (1 - e^{-\tau})/\tau$ derived from an expanding sphere geometry. We used the CO collisional data from the LAMDA database \citep{Schoier2005}. With an MCMC approach following \cite{Yang2017}, we explored the parameter space consisting of the kinetic temperature of the molecular gas (T$_{\mathrm{K}}$), the volume density (n$_{\mathrm{H_2}}$), the column density of CO per unit velocity gradient (N$_{\mathrm{CO}}$/dV), and the solid angle ($\Omega_{\mathrm{app}}$) of the source. The overall shape of the CO SLEDs only depends on T$_{\mathrm{K}}$, n$_{\mathrm{H_2}}$ and N$_{\mathrm{CO}}$/dV, and scales with $\Omega_{\mathrm{app}}$ (the magnification factors are also included in this factor). Therefore, we only focus on the parameters T$_{\mathrm{K}}$, n$_{\mathrm{H_2}}$ and N$_{\mathrm{CO}}$/dV hereafter. As noted in the previous section, the CO(6$-$5) flux of HerBS-43b is quite uncertain and is an outlier when compared to the other transitions, therefore we chose to exclude it from our SLED modelling analysis. We note that a detailed analysis for the SLED of HerBS-89a was presented in \cite{Berta2021}, where additional higher-$J$ transitions were included. Here we present the analysis of the rest of the sample.

As most of the transitions of the CO lines observed are below $J_{up}=$6, there will be insufficient data to constrain the highly-excited molecular gas component, assuming the HerBS galaxies are similar to those high-$z$ SMGs, in which the CO excitation is dominated by two components peaking around $J_{up}=$6 and $J_{up}=$8 respectively \citep[][]{Yang2017,Canameras2018}. To better constrain the posteriors, we have given slightly tighter boundaries for the flat priors of n$_{\mathrm{H_2}}$ and N$_{\mathrm{CO}}$/dV than those used in \cite{Yang2017} (while other priors stay the same). Taking the values of the parameters from statistically studied SMG samples \citep[][]{Yang2017,Canameras2018}, we have chosen flat priors of $\rm log(n_{\mathrm{H_2}}/cm^{-2})=2.0-5.5$ and $\rm log(N_{\mathrm{CO}}/dV/cm^{-2} (km/s)^{-1}) = 15.5-18.5$. Similarly, we also limited the range of the thermal pressure P$_\mathrm{thermal}$ (defined by P$_\mathrm{thermal} = n_{\mathrm{H_2}} \times T_{\mathrm{K}}$) to be within 10$^4$ and 10$^7$\,K\, cm$^{-3}$. 

Two hundred walkers have been deployed with five hundred iterations after the one hundred burn-in ones. Thus, a total of 100,000 points of the solutions have been explored in parameter space. 

The results of the radiative transfer analysis are reported in Table~\ref{tab:sledfit}, indicating the $\pm1\sigma$ values and the median of the posteriors. The maximum likelihood values are also listed in the table. 
The lack of measurements at $J_{up}>$6 for the majority of our sample means that the CO SLED models within $\pm1\sigma$ of the best fit are significantly different at high-$J$, leading to a relatively flat posterior of T$_{\mathrm{K}}$, and therefore a poor constraint on the maximum likelihood values (max). Therefore, we use the median (med) values to describe the fits. The CO data can be described with models of dense warm gas of $n_{\rm H_2} = 0.31-8.5\times10^3$\,cm$^{-3}$, and $T_{\rm K} = 100-200$\,K. These values are in general agreement with what has been found previously in high-$z$ luminous DSFGs and SMGs \citep[e.g.][]{Combes2012,Danielson2013,Riechers2013,Spilker2014,Yang2017,Harrington2021}.

\begin{figure}
\centering
\includegraphics[width=0.55\textwidth]{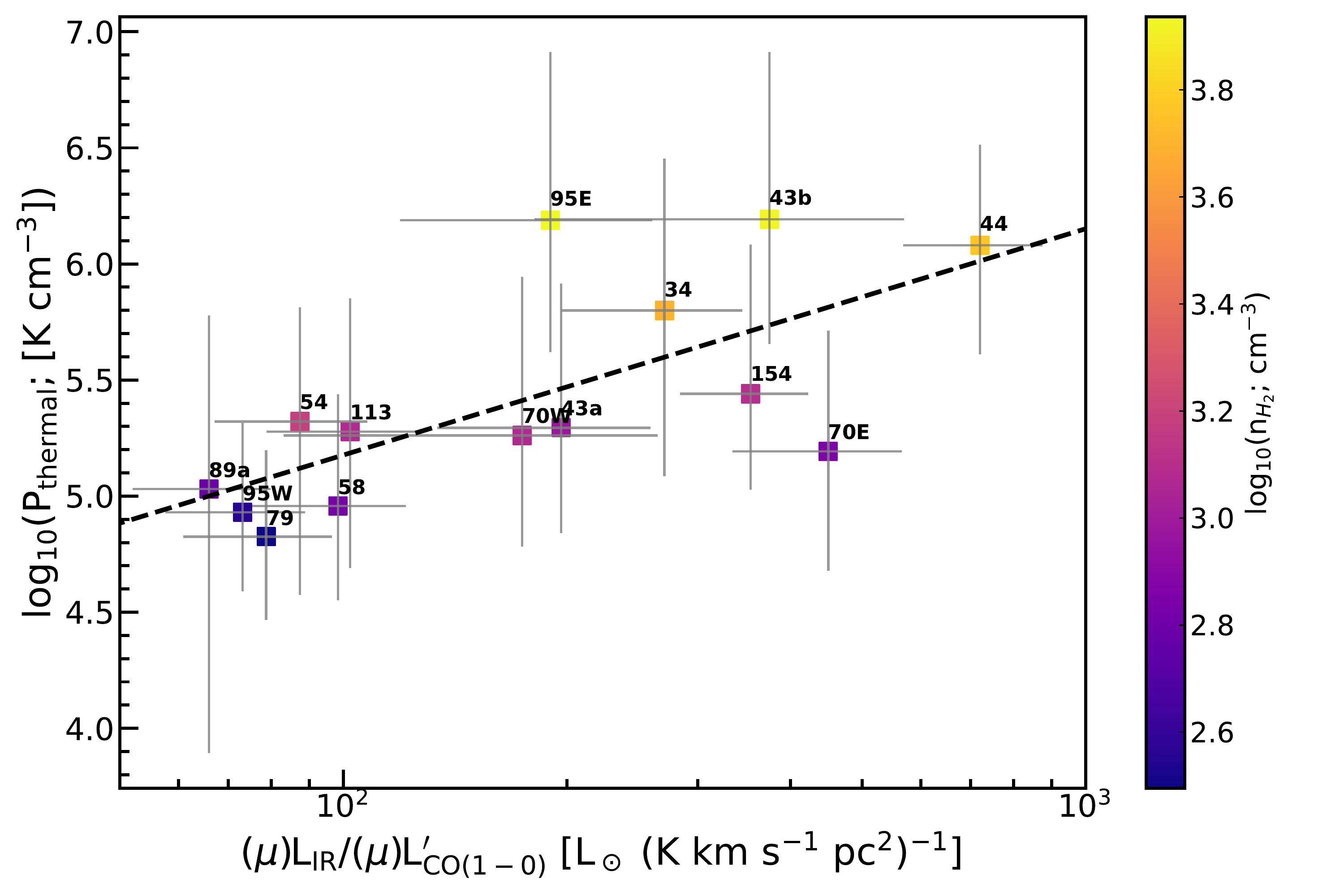}\\
\caption{Median gas pressure (P$_{\rm thermal}$) from the median of the fitted models of the SLEDs, as a function of the L$_{\rm IR}$/L$^\prime_\mathrm{CO(1-0)}$. The sources are plotted with gradient colour that corresponds to the median volume densities ($n_\mathrm{H_2}$).
} \label{fig:pthermal}
\end{figure}

A relation between $\rm P_\mathrm{thermal}$ and SFE has been discussed in theoretical work \citep[e.g.][]{Elmegreen1997,Wong2002}, suggesting that the main driver of the SFE in collapsing molecular clouds is the thermal gas pressure, or alternatively that higher pressures lead to more molecular clouds and a higher molecular gas fraction. \cite{Yang2017} examined whether there is evidence for this through the relationship between $\rm P_\mathrm{thermal}$ and L$_{\rm IR}$/L$^\prime_\mathrm{CO(1-0)}$ (as a proxy for SFE), in a sample of $Herschel$ selected luminous lensed DSFGs. A statistically significant correlation was found between $\rm log_{10}(P_{thermal})$ and $\rm log_{10}(L_{\rm IR}/L^\prime_\mathrm{CO(1-0)})$, spanning the range of environments from nearby galaxies to high-$z$ luminous DSFGs, with no evidence of it being driven by the $n_{\rm H_2}$ or $T_{\mathrm{K}}$ values. Following \cite{Yang2017}, in Figure~\ref{fig:pthermal} we plot the median values of $\rm P_\mathrm{thermal}$ from the radiative transfer analysis as a function of L$_{\rm IR}$/L$^\prime_\mathrm{CO(1-0)}$. In agreement with what was reported in \cite{Yang2017} we find that the data are following a correlation. We fit a line through the data in log-space and find a strong correlation of $\rm log_{10}(P_{thermal}) \varpropto 0.97\times log_{10}(L_{\rm IR}/L^\prime_\mathrm{CO(1-0)})$, with Pearson coefficient and p-value of 0.69 and 0.006, respectively. In addition, we color the datapoints based on the $\rm log_{10}(n_{H_2})$ value, which allows us to simultaneously examine the presence of a dependence of $\rm n_{H_2}$ with L$_{\rm IR}$/L$^\prime_\mathrm{CO(1-0)}$. We observe a trend in the $\rm n_{H_2}$ values showing an increase with higher L$_{\rm IR}$/L$^\prime_\mathrm{CO(1-0)}$. To examine wether this positive trend of $\rm n_{H_2}$ values could be the driver of the observed correlation between $\rm P_\mathrm{thermal}$ and L$_{\rm IR}$/L$^\prime_\mathrm{CO(1-0)}$, we calculate the Pearson coefficient and p-value for a correlation between $\rm n_{H_2}$ and L$_{\rm IR}$/L$^\prime_\mathrm{CO(1-0)}$. Indeed, we find evidence for a correlation of $\rm n_{H_2}$ with L$_{\rm IR}$/L$^\prime_\mathrm{CO(1-0)}$, with a Pearson coefficient and p-value of 0.63 and 0.016, respectively. No evidence of a correlation between $T_{\mathrm{K}}$ and L$_{\rm IR}$/L$^\prime_\mathrm{CO(1-0)}$ is found. This suggests that the $\rm n_{H_2}$ trend could be significantly contributing to the observed relation, but is not likely the driver as the correlation of $\rm n_{H_2}$ is less significant than that of $\rm P_\mathrm{thermal}$. However, we remind the reader that our radiative transfer analysis is limited by the lack of CO transitions at $J > 6$.

\section{Conclusions}\label{sec:conclusions}
In this paper, we have presented \aco\ observations using the VLA for 14 luminous DSFGs, including two binary systems, in the redshift range $2 < z < 4$. These sources are part of the pilot project sample for the large NOEMA project $z$-GAL (Cox et al. in prep.), and were originally presented in \citet{Neri2020}. Our main findings are the following: 

\begin{itemize} 
    \item We successfully detected the \aco\ emission in the entirety of our sample, resolving the molecular emission into extended and complex morphologies in nearly half the sample (7/14 sources). For the majority of our sample, the \aco\ line profiles agree with those of the higher-$J$ transitions. 
    \item Four of the sources were detected in the underlying radio continuum at observed frequencies of 20-38\,GHz. Simple SED fitting to the far-infrared and radio emission reveals that HerBS-43a, -43b, and -54 have radio emission consistent with their SFRs, while HerBS-70E has additional synchrotron radiation from an AGN.  
    \item We find that our sources have CO luminosities of $(\mu)L^\prime_{CO} = 0.4-2.9 \times 10^{11}$\,K\,km\,s$^{-1}$\,pc$^2$. From these \aco\ luminosities, we estimated molecular gas masses of $(\mu)M_{\rm H_2} = 1.3-8.6\times10^{11}$\,M$_\odot$. By combining our \aco\ luminosities with the total infrared (8--1000$\mu$m) luminosities from \citet{Neri2020}, we compare the molecular gas depletion times ( $t_{\rm dep}$) and star formation efficiencies (SFEs) to other galaxy samples. We find that our sources have SFEs consistent with high-$z$ DSFGs and low-$z$ ULIRGs. Similarly, the depletion times of our sources cover a wide range and are consistent with both the main sequence and the starburst phase, with values of $t_{\rm dep} = 40-460$\,Myr.
    \item We find a wide range of gas to dust ratios, $\rm M_{\rm H_2}/M_{\rm dust} = 27 - 167$, in agreement with previous measurements for the majority of our sample. Three sources have surprisingly low ratios of $<40$, which could be a result of variation of $\alpha_{CO}$ amongst the sample, or due to the presence of an AGN.
    \item We find CO temperature brightness ratios of $r_{32/10} = 0.4-1.4$, $r_{43/10} = 0.4-1.7$, and $r_{54/10} = 0.3-1.3$, with median values of 0.7, 0.6, and 0.7 respectively. The ratios are relatively consistent for the different transitions, with similar distributions. 
    The range of values observed are in agreement with previous observations of high-$z$ DSFGs and AGN \citep[e.g.][]{Sharon2016}. 
    \item We find a wide range in the shapes of the CO SLEDs of our sample, highlighting the importance of \aco\ in revealing the range of excitation in such galaxies. We compare the SLEDs of our sources to those of main sequence, typical SMGs, and starburst-quasar systems. The majority of our sample is consistent with high-$z$ SMGs, with the exception of HerBS-43b, HerBS-44, and HerBS-95E that are more  comparable with the quasar-starburst systems, Cloverleaf and APM08279$+$5255. 
	\item Finally, we perform radiative transfer modeling of the SLEDs following \cite{Yang2017}. We find that our sample can be described with models of $\rm n_{H_2} = 0.3-8.5\times10^3$\,cm$^{-3}$ and temperatures of $\rm T_K = 100 - 200\,K$. However, the results of the modelling are limited due to the lack of CO observations at the highest transitions. We examine the possible relation of the thermal gas pressure with SFE, by using the approximation of the $\rm L_{IR}/L^\prime_{CO(1-0)}$. We find strong evidence of a correlation, in agreement with the theoretical idea of thermal gas pressure having a direct role in the star formation process of these galaxies.

\end{itemize}

The results of this study emphasize the importance of anchoring the CO SLED to the ground state in order to determine the properties of the dense gas. Furthermore, by probing the underlying radio continuum we can trace the contribution to the radio emission from AGN and/or intense star formation in high-$z$ DSFGs.
Building on the previous studies of \aco\ in high-$z$ DSFGs, our successful detection of the full sample with the high angular resolution possible with the VLA, revealed a wide range of galaxy and excitation properties for a sample that would otherwise be classified as similarly luminous highly star-forming systems. The demonstrated success of this Pilot study highlights the progress we will be able to achieve by measuring CO(1-0) in the full $z$-GAL sample of 126 {\it Herschel} selected sources, in understanding the range of physical conditions that lead to the formation and evolution of luminous high-$z$ DSFGs.
\newpage
\begin{acknowledgements}
      We thank the anonymous referee for their comments that helped to improve the clarity of various aspects of the paper.  This work is based on observations carried out under program VLA/20A-083 using the National Radio Astronomy Observatory's (NRAO)  Karl G. Jansky Very Large Array (VLA). The National Radio Astronomy Observatory is a facility of the National Science Foundation operated under cooperative agreement by Associated Universities, Inc. This work benefited from the support of the project Z-GAL ANR-AAPG2019 of the French National Research Agency (ANR). B.M.J. acknowledges the support of a studentship grant from the UK Science and Technology Facilities Council (STFC) and funding from the Deutsche Forschungsgemeinschaft (DFG, German Research Foundation) –Project SCHI 561/3-1. C.Y. acknowledges support from an ESO Fellowship. T.B. acknowledges funding from NAOJ ALMA Scientific Research Grant 2018-09B and JSPS KAKENHI No.~17H06130. HD acknowledges financial support from the Agencia Estatal de Investigaci\'on del Ministerio de Ciencia e Innovación (AEI-MCINN) under grant (La evoluci\'on de los c\'iumulos de galaxias desde el amanecer hasta el mediod\'ia c\'osmico) with reference (PID2019-105776GB-I00/DOI:10.13039/501100011033) and acknowledge support from the ACIISI, Consejer\'ia de Econom\'ia, Conocimiento y Empleo del Gobierno de Canarias and the European Regional Development Fund (ERDF) under grant with reference PROID2020010107. S.J. acknowledges the financial support from the European Union's Horizon research and innovation program under the Marie Sk\l{}odowska-Curie grant agreement No. 101060888.
\end{acknowledgements}

%BIBLIOGRAPHY
\bibliography{vla_co.bib}
\bibliographystyle{aasjournal}
%%%%%%%%%%%%%

\begin{appendix}\label{Appendix}
\section{Individual sources}\label{app:ind_sources}
In this section, we provide details on the \aco\ emission line profiles and the integrated-velocity maps for each source and compare them with the NOEMA 3 and 2-mm results described in \citet{Neri2020}.

{\bf HerBS-34} is well detected in the \aco\ emission line displaying a profile that is comparable to the CO(3$-$2) line, although the ratio of  \aco\ FWHM over the $\Delta$V of higher-$J$ transitions 
(FWHM/$\Delta$V) is 1.8$\pm$0.3. The large ratio is likely due to the lower SNR of the \aco\ line compared to the higher-$J$ transitions. 
The CO emission remains fairly compact even when resolved, with an estimated size (deconvolved from the beam) of $1\farcs7\times0\farcs9$ ($14\times7$\,kpc).

{\bf HerBS-43a} displays a similarly broad \aco\ emission line as the higher-$J$ CO lines, with a ratio of  FWHM/$\Delta$V$=1.1\pm0.2$. 
The source is partially resolved, with an estimated size (deconvolved from the beam) of $1\farcs2\times0\farcs9$ ($9\times7$\,kpc).

{\bf HerBS-43b},is the highest redshift source in our sample, and therefore has the weakest \aco\ emission line of the entire sample with a line flux of $\rm 0.064 \pm 0.033 \, Jy \, km s^{-1}$. The line profile of the \aco\ is nearly identical to that of the CO(4$-$3) line, showing the same double peaked profile. The ratio of FWHM/$\Delta$V$=0.9\pm0.4$.
 The source is partially resolved, with an estimated size (deconvolved from the beam) of $1\farcs1\times0\farcs3$ ($8\times2$\,kpc). 

{\bf HerBS-44} has a strong \aco\ emission line with a single peaked profile and a $\rm FWHM = 377 \pm 64 \, km \, s^{-1}$ that is somewhat narrower than the high-$J$ transitions, with a ratio of FWHM/$\Delta$V$=0.7\pm0.1$. The source is elongated in the east-west direction with an estimated size (deconvolved from the beam) of $1\farcs1\times0\farcs5$ ($9\times4$\,kpc). 
          
{\bf HerBS-54} has a very broad \aco\ line with FWHM$ = 1087 \pm 176 \, \rm km \, s^{-1}$ comparable to the high-$J$ CO lines, with a ratio of FWHM/$\Delta$V$=1.1\pm0.3$. With the new VLA observations, the source is resolved into a complex morphology showing two peaks within extended weaker emission. The peaks are separated by $\sim1\farcs0$ and the \aco\ emission extends over a region of $\sim2\farcs2$ (18\,kpc).
     
{\bf HerBS-58} shows a complex morphology with two peaks along an arc-like feature to the west and a well-defined peak to the east. The overall extent of the CO emission is estimated to be $2\farcs1\times1\farcs6$ ($18\times14$\,kpc). The \aco\ line profile, which is best described by a single and relatively narrow gaussian of $\rm FWHM = 363 \pm 64 \, km \, s^{-1}$, is red-shifted at $\rm +300 \, km \, s^{-1}$ with respect to the center velocity corresponding to a $z_{spec}=2.0842$ as reported in \citet{Neri2020}. The spectroscopic redshift for HerBS-58 was derived from the double-peaked $\rm [CI](^3P_1-^3P_0)$ emission line where both peaks have similar line fluxes. It is interesting to note that the $\rm CO$(3$-$2) emission line shows a strong red-shifted peak (at $\rm \sim 300 \, km \, s^{-1}$), corresponding to the \aco\ emission line, whereas the blue-shifted component is at least $2 \times$ weaker. There is no clear evidence for a blue-shifted CO component in the VLA data. A more detailed study of this source based on higher angular resolution NOEMA data will be presented in Ismael et al. (in prep.)

{\bf HerBS-70} is a binary system which was originally resolved in \citet{Neri2020}, including the sources HerBS-70E and HerBS-70W that are separated by $16\farcs5$ corresponding to a projected distance of $\rm \sim 140 \, kpc$.
Both are well detected and resolved in the \aco\ emission. {\bf HerBS-70E} (to the east) has a wide emission line with $\rm FWHM = 622 \pm 130$\,km\,s$^{-1}$ comparable to the higher-$J$ transitions, with a ratio of FWHM/$\Delta$V$=0.8\pm0.2$. In contrast, {\bf HerBS-70W}  has narrow line of $\rm FWHM = 197\pm 39$\,km\,s$^{-1}$, similar to the one seen in the higher-$J$ CO lines with a ratio of FWHM/$\Delta$V$=1.4\pm0.3$. Both sources display extended and somewhat resolved \aco\ emission with sizes of $1\farcs6\times1\farcs2$ ($13\times10$\,kpc), and $2\farcs1\times1\farcs1$ ($18\times9$\,kpc), for HerBS-70E and HerBS-70W, respectively.

{\bf HerBS-79} displays a broad \aco\ emission line with $\rm FWHM = 787 \pm 125$\,km\,s$^{-1}$ that is comparable to the CO(3$-$2) and CO(4$-$3) profiles, with a ratio of FWHM/$\Delta$V$=0.9\pm0.2$. The moment-0 map shows an extended arc-like morphology with a size of $2\farcs9\times1\farcs0$ ($25\times8$\,kpc) in the east-west direction. 

The VLA data of {\bf HerBS-89a} were originally published separately in \citet{Berta2021}, a dedicated study of the source that combined the \aco\ observations with high angular resolution NOEMA observations of the $\rm CO(9-8)$ emission line and additional emission or absorption lines of molecular tracers (including $\rm H_2O$ and $\rm OH^+$). We have reanalysed the \aco\ emission to be consistent with the analysis of the whole sample. We find a similar morphology and line profile as reported in \citet{Berta2021}. However, the \aco\ from our extraction is slightly wider, with a $\rm FWHM = 1586 \pm 244$\,km\,s$^{-1}$ (versus $1433\pm293$\,km\,s$^{-1}$) and the integrated line flux is significantly larger with $I_{\rm CO(1-0)} = 1.08\pm0.22$Jy\,km\,s$^{-1}$ (versus $0.64\pm0.13$Jy\,km\,s$^{-1}$). The difference between the two results is due to the larger extraction region used in our analysis. The morphology of the \aco\ emission is arc-like, which was clearly resolved in the higher angular resolution NOEMA observations into a partial $1\farcs0$ diameter Einstein ring in the dust emission and the molecular emission lines of $\rm CO(9-8)$ and $\rm H_2O(2_{02}-1_{11})$. Based on a lensing model, the magnification of HerBS-89a was estimated to be $\mu \sim 5$ \citep[see][]{Berta2021}.

{\bf HerBS-95} is the second binary system resolved by \citet{Neri2020}, including the sources HerBS-95E and HerBS-95W, which are separated by $16\farcs4$ corresponding to a projected distance of $\rm \sim 130 \, kpc$. {\bf HerBS-95E} displays a \aco\ emission line that is somewhat blue-shifted with respect to the expected rest-frame zero velocity, with a ratio of FWHM/$\Delta$V$=0.8\pm0.2$. The emission is slightly extended with a size of $1\farcs5\times0\farcs8$ ($12\times6$\,kpc). {\bf HerBS-95W} shows a double-peaked line profile with a $\rm FWHM = 522$\,km\,s$^{-1}$, in agreement with the higher-$J$ transitions, with a ratio of FWHM/$\Delta$V$=1.0\pm0.2$. The \aco\ emission has a centrally-peaked but extended structure with a size of $2\farcs5 \times 1\farcs9$ ($20\times15$\,kpc). 

{\bf HerBS-113} has a \aco\ profile somewhat narrower than the higher-$J$ transitions with a $\rm FWHM = 497$\,km\,s$^{-1}$, with a ratio of FWHM/$\Delta$V$=0.6\pm0.2$. The morphology of \aco\ emission is complex with three distinct peaks, distributed along an elongated ring-like structure and extending over $\sim3\arcsec$ (24\,kpc). This morphology has been confirmed by higher resolution CO observations that will be presented in Ismail et al. (in prep.).   

{\bf HerBS-154} displays a strong single-peaked \aco\ emission line with a $\rm FWHM = 384 \pm 54$\,km\,s$^{-1}$, comparable to the higher-$J$ CO and $\rm [CI](^3P_1-^3P_0)$ lines, with a ratio of FWHM/$\Delta$V$=1.2\pm0.2$. The \aco\ emission has an extended arc-like morphology with a size of $2\farcs7\times1\farcs7$ ($20\times13$\,kpc).

\section{Lensing}\label{app:lensing}

The large sub-millimeter surveys made with telescopes such as {\it Herschel} or the South Pole Telescope, have revealed a significant population of bright high-redshift SMGs that are strongly gravitionally lensed, 
 that becomes dominant for high flux densities ($S_{\rm 500 \mu} > 100 \rm mJy$) \citep[e.g.,][]{Negrello2010, Wardlow2013,Mocanu2013}. We therefore expect that some sources in our sample will be lensed, as discussed in \cite{Neri2020}. The new \aco\ data have revealed that sources HerBS-54, HerBS-58, HerBS-79, HerBS-89a, HerBS-113 and HerBS-154, have complex morphologies such as arc-like structures, filaments or multiple components,  suggesting that these sources could be gravitationally lensed (see Fig. 1). In the case of HerBS-89a higher resolution followup observations confirmed that it is indeed lensed \citep[see][]{Berta2021}. However, for the rest of the sources, a derivation of the magnification based on the available data remains uncertain and deeper observations with higher angular resolution are needed to make a detailed lensing model. 

In past studies, the relationship between $L^\prime_{\rm CO(1-0)}$ and the full width half maximum (FWHM) of the \aco\ emission line has been discussed as a tool to estimate the magnification factor of high-redshift galaxies \cite[e.g.,][]{Harris2012, Bothwell2013, Aravena2016, Neri2020}. As $L^\prime_{\rm CO(1-0)}$ is a measure of $M_{\rm H_2}$ and the \aco\ FWHM  is related to the dynamical mass, this relationship is seen as a proxy to the Tully-Fisher relation \cite[e.g.,][]{Isbell2018}. Therefore, any sources lying above the expected correlation would be lensed. However, as shown in \citet{Aravena2016}, the use of the $L^\prime_{\rm CO(1-0)}$ versus FWHM relationship for measuring magnifications is highly unreliable, and can be typically off by factors of $\sim2$ or more.  

\begin{figure}
\centering
\includegraphics[width=0.85\textwidth]{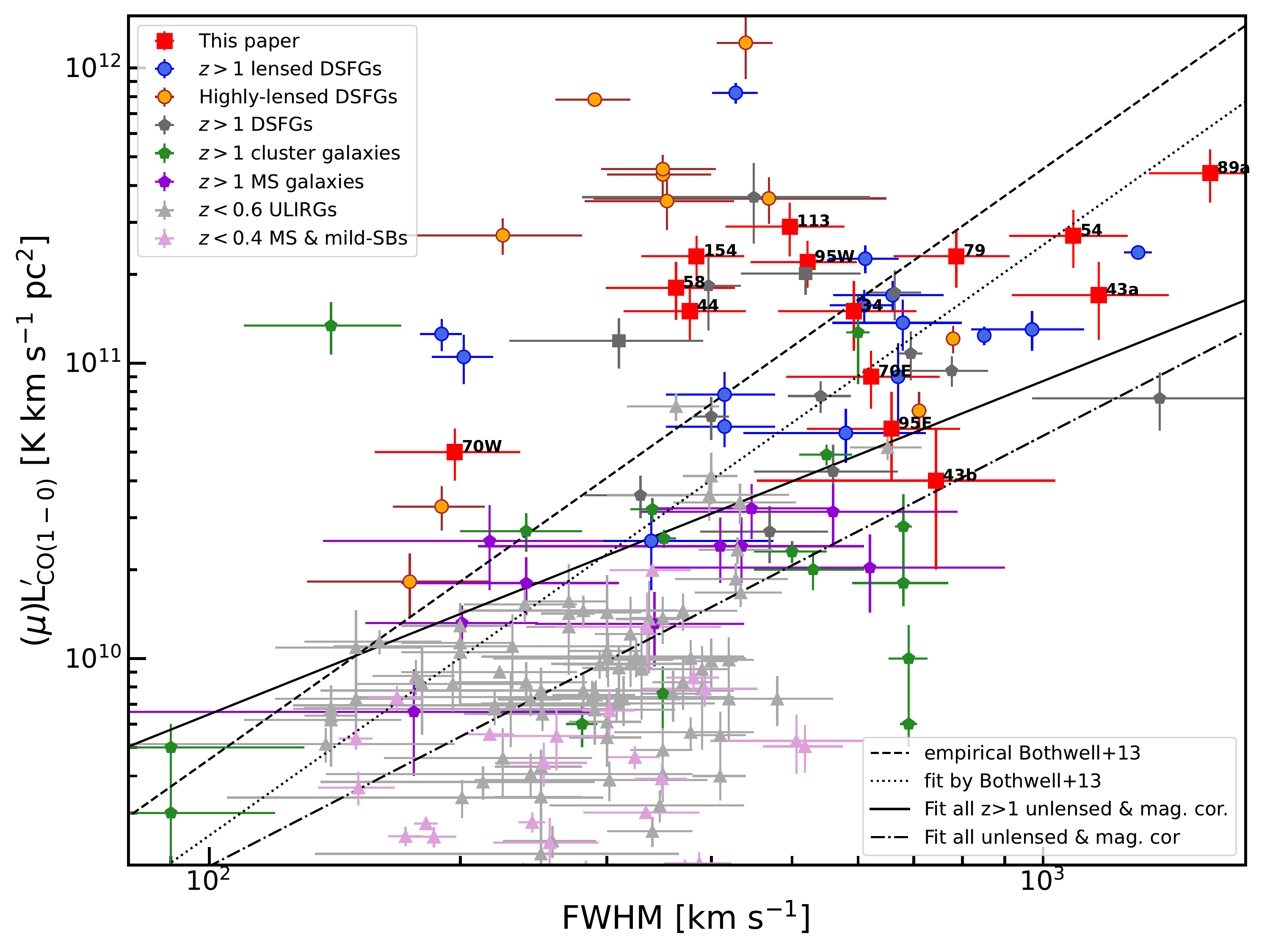}
\caption{L$^{\prime}_\mathrm{CO(1-0)}$ versus the full width half max (FWHM) of the $\rm CO(1-0)$ emission line. The sample of sources presented in this paper are shown with large red squares (and identified by their name), and are compared to other high-$z$ lensed and unlensed luminous DSFGs \citep[][and references therein]{Ivison2011, Harris2012, Aravena2016,  Bakx2020b, Carilli-Walter2013}, $z>1$ main-sequence (MS) galaxies \citep{Aravena2014} and cluster galaxies \citep{Rudnick2017, Wang2018}, as well as low-$z$ ultra-luminous infrared galaxies (ULIRGs), mild-starburst (mild-SBs) and MS galaxies \citep{Solomon1997, Chung2009, Combes2011, Geach2011, Villanueva2017}. Only sources with reported $\rm CO(1-0)$ emission lines have been included to avoid uncertainties in estimating L$^{\prime}_\mathrm{CO(1-0)}$ from higher-$J$ transitions. No correction for magnification was applied to the CO luminosities. Sources with magnifications $>10$ are classified as Highly-lensed SMGs. The dotted and dashed lines represent the relationships described in \cite{Bothwell2013} (see text and Eq.~\ref{eq:L'co}); the solid and dash-dotted lines display fits to all the $z>1$ galaxies and the entire sample of sources, respectively, that are unlensed and have been corrected for magnification. 
}
\label{fig:lcodv}
\end{figure}

In Figure~\ref{fig:lcodv}, we plot $L^\prime_{\rm CO(1-0)}$ as a function of the \aco\ FWHM for the sources of our sample together with sources previously observed in \aco. We limit our comparison to high-redshift sources that are unlensed and sources for which the gravitational magnification has been reliably estimated based on lensing model analysis. Therefore, we remove uncertainties that can accompany conversions from higher-$J$ CO transitions or sources which were defined as magnified using this same relation. The known lensed sources (identified in the figure) were not corrected for magnification.  
The dotted line shows the best-fitting relationship from \cite{Bothwell2013} and the dashed line represents the derived parametrization for $L^\prime_{\rm CO}$ by \cite{Bothwell2013}  for a disk galaxy, that can be described as:
\begin{equation}\label{eq:L'co}
{L^\prime_{\rm CO}} = C \, \left (\frac{\Delta V}{2.35} \right)^2 \frac{R}{\alpha \, G},
\end{equation}
where $\Delta V$ is the FWHM of the CO in $\rm km \, s^{-1}$, $R = 7 \, \rm kpc$ is the radius of the CO emitting region in parsecs, 
$G$ is gravitational constant, $\alpha$ is the CO luminosity to gas mass conversion factor, 
and $C=2.1$ is the constant parametrizing the galaxy's disk morphology. 

We use the compilation of \aco\ detected galaxies and fit all sources that are unlensed and the lensed sources after correcting for magnification, as well as only those at $z>1$. Both relationships show a less steep correlation than previously found. We note that we plot the lensed sources without the magnification correction, even though we correct them for magnification for the fit, in order to better visualise the parameter space covered by the lensed sources, in relation to the unlensed sources and the relations.

Overall, we find a large scatter for both the lensed and unlensed galaxies, consistent with what has been reported in previous studies \citep[e.g.,][]{Aravena2016}. Sources with magnifications of $<10$ mostly cover a similar region of the parameter space as the unlensed sources, with only highly lensed galaxies ($\mu>10$) showing clear offsets. This strengthens the conclusions of \citet{Aravena2016} on the unreliability of this relation as a precise measure of magnification \citep[see also discussions in, e.g.,][]{Dannerbauer2017,Aravena2019,Jin2021}. It is interesting to note the specific example of HerBS-89a, which based on the $L^\prime_{\rm CO(1-0)}$--FWHM relationship would be considered to be an unlensed source when comparing to the \cite{Bothwell2013} relation, and therefore a HyLIRG \citep[see][]{Neri2020}, was found to be lensed with a magnification of $\sim 5$ based on high-angular resolution observations and a lensing model \citep{Berta2021}. For the above reasons, we do not attempt to determine the magnifications of our sample using this method. 

\end{appendix}

\end{document}